\documentclass[manuscript,screen]{acmart}

\usepackage{amsmath,amssymb,amsfonts}
\usepackage{algorithmic}
\usepackage{graphicx}
\usepackage{textcomp}
\usepackage{xcolor}
\usepackage{subfigure}
\usepackage{multirow}
\usepackage{framed}
\usepackage{pifont}
\usepackage{fancyhdr}
\usepackage[framemethod=TikZ]{mdframed} 

\usepackage{multirow}
\usepackage{amssymb}

\begin{document}


\pagestyle{fancy}
\fancyhf[RE]{} 
\fancyhead[RE]{S. Zhang et al.}
\newcommand{\setParDis}{\setlength {\parskip} {0.1cm} }

\title{Empowering Agile-Based Generative Software Development through Human-AI Teamwork}

\author{Sai Zhang}\thanks{S. Zhang also with CSIRO's Data61}
\email{zhang_sai@tju.edu.cn}
\orcid{0000-0001-8972-2824}
\affiliation{
    \institution{College of Intelligence and Computing, Tianjin University}
    \country{China}
    \postcode{300072}
  }

\author{Zhenchang Xing}
\email{zhenchang.xing@data61.csiro.au}
\orcid{https://orcid.org/0000-0001-7663-1421}
\affiliation{%
  \institution{CSIRO’s Data61 \& Australian National University}
  \country{Australia}
  \postcode{0200-2600-2601}
}

\author{Ronghui Guo}
\email{ronghui_guo@tju.edu.cn}
\orcid{https://orcid.org/0000-0002-1586-7040}
\author{Fangzhou Xu}
\email{xu_fangzhou@tju.edu.cn}
\orcid{https://orcid.org/0009-0000-0667-4653}  
\author{Lei Chen}\thanks{This work was supported by the Project of Science and Technology Research and Development Plan of China Railway Corporation (N2023J044).}
\email{2022244117@tju.edu.cn}
\orcid{https://orcid.org/0009-0006-8455-7253}
\author{Zhaoyuan Zhang}
\email{zhaoyuanzhang@tju.edu.cn}
\orcid{https://orcid.org/0009-0000-9183-1173}
\author{Xiaowang Zhang$^*$}
\email{xiaowangzhang@tju.edu.cn}
\orcid{https://orcid.org/0000-0002-3931-3886}
\thanks{*Corresponding author}
\author{Zhiyong Feng}
\email{zyfeng@tju.edu.cn}
\orcid{https://orcid.org/0000-0001-8158-7453}
\author{Zhiqiang Zhuang}
\email{zhuang@tju.edu.cn}
\orcid{https://orcid.org/0000-0003-0081-1703}
\affiliation{%
  \institution{College of Intelligence and Computing, Tianjin University}
  \country{China}}

\begin{abstract}
In software development, the raw requirements proposed by users are frequently incomplete, which impedes the complete implementation of software functionalities. With the emergence of large language models, the exploration of generating software through user requirements has attracted attention. Recent methods with the top-down waterfall model employ a questioning approach for requirement completion, attempting to explore further user requirements. However, users, constrained by their domain knowledge, result in a lack of effective acceptance criteria during the requirement completion, failing to fully capture the implicit needs of the user. Moreover, the cumulative errors of the waterfall model can lead to discrepancies between the generated code and user requirements. The Agile methodologies reduce cumulative errors of the waterfall model through lightweight iteration and collaboration with users, but the challenge lies in ensuring semantic consistency between user requirements and the code generated by the agent. To address these challenges, we propose AgileGen, an agile-based generative software development through human-AI teamwork. Unlike existing questioning agents, AgileGen adopts a novel collaborative approach that breaks free from the constraints of domain knowledge by initiating the end-user perspective to complete the acceptance criteria. By introducing the Gherkin language, AgileGen attempts for the first time to use testable requirement descriptions as a bridge for semantic consistency between requirements and code, aiming to ensure that software products meet actual user requirements by defining user scenarios that include acceptance criteria. Additionally, we innovate in the human-AI teamwork model, allowing users to participate in decision-making processes they do well and significantly enhancing the completeness of software functionality. To ensure semantic consistency between requirements and generated code, we derive consistency factors from Gherkin to drive the subsequent software code generation. Finally, to improve the reliability of user scenarios, we also introduce a memory pool mechanism, collecting user decision-making scenarios and recommending them to new users with similar requirements. AgileGen, as a user-friendly interactive system, significantly outperformed existing best methods by 16.4\% and garnered higher user satisfaction.

\end{abstract}



\maketitle

\section{Introduction}

Generative software development,  based on the capability of automated code generation~\cite{DBLP:journals/computers/PaoloneMPF20[1],DBLP:journals/corr/abs-2303-17568[2],DBLP:conf/sigsoft/ShenZDGZL22[3],DBLP:conf/kbse/WangLZLLWC22[4],DBLP:journals/corr/abs-2303-17780[5]}, involves customizing personalized software according to user requirements. It is widely recognized that realizing generative software development requires overcoming two major challenges: requirement analysis and software design~\cite{10.1145/3487569}. Deep learning technology is applied at various stages of generative software development~\cite{DBLP:conf/emnlp/0034WJH21[19],DBLP:conf/emnlp/FengGTDFGS0LJZ20[20],DBLP:journals/corr/abs-2107-03374[21]} to reduce costs and enhance efficiency.

\begin{figure}[h]
	\centering
	\includegraphics[width=1\textwidth]{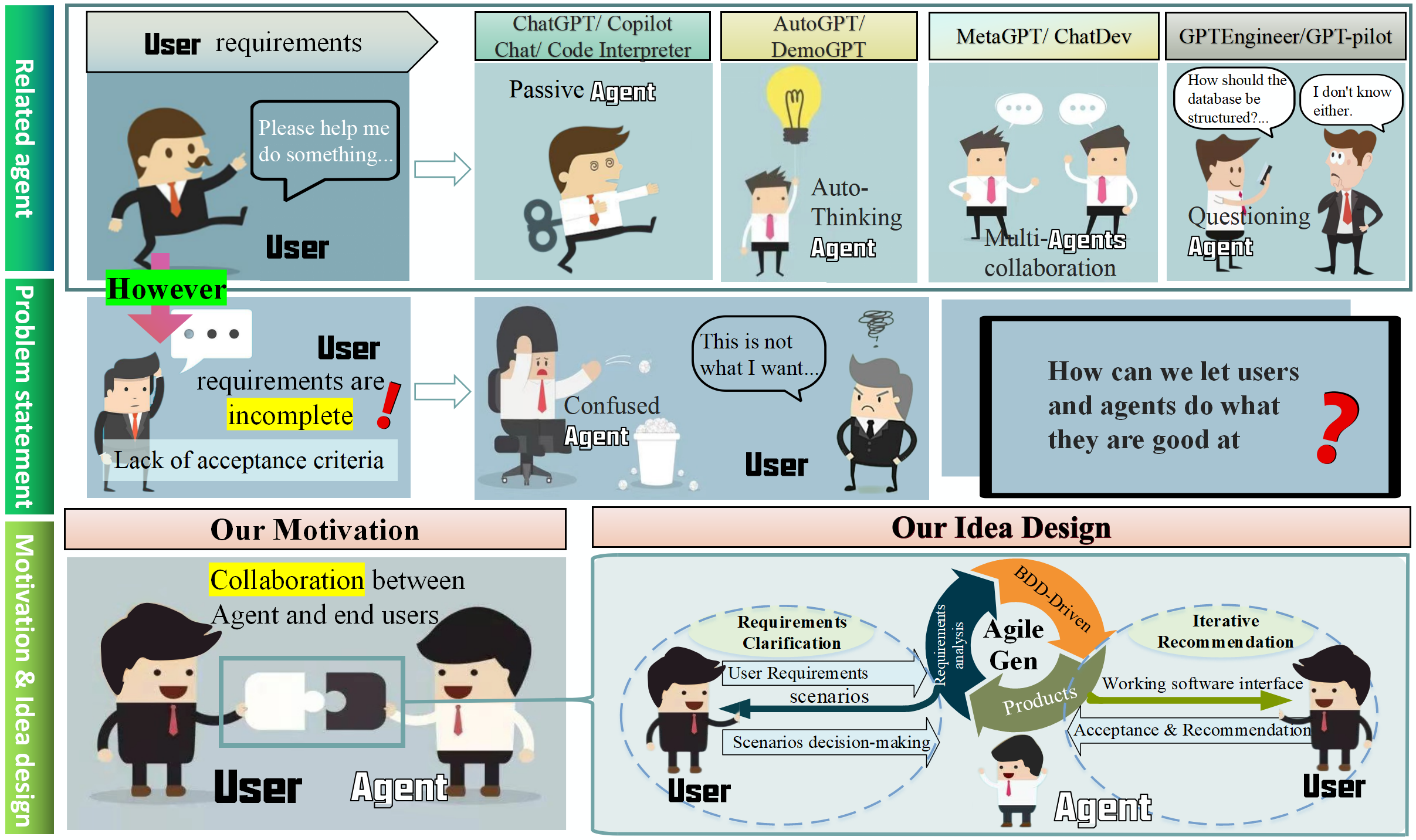}
	\caption{\label{fig:fig0} Problem Statement Diagram. Users are unsure how to drive the Agent to generate desired software, and the Agent does not know how to fulfill user requirements. We have built a bridge between users and the Agent, facilitating collaboration between human decision-making skills and the Agent's coding capabilities. This collaboration has created a generative software development Agent with lightweight iterative feedback.}
\end{figure}

Nowadays, Large Language Models have made significant progress in automated code generation~\cite{10403378,10298349}, making generative software development increasingly mainstream~\cite{DBLP:journals/corr/abs-2307-07924[1]2,DBLP:journals/corr/abs-2308-00352[2]2}. These models take user requirements as input and software code as output. As shown in Figure~\ref{fig:fig0}, existing generative software development intelligent agents are divided into four categories according to their design philosophy:
\textbf{(1) Passive Agent} (e.g., ChatGPT~\cite{DBLP:journals/corr/abs-2302-04023[11]}, Copilot Chat\footnote{https://docs.github.com/en/copilot/github-copilot-chat/about-github-copilot-chat}, Code Interpreter\footnote{https://openai.com/blog/chatgpt-plugins}): These agents directly execute any commands they receive from users. While they are good copilots for developers, they rely more on step-by-step guidance from users and cannot analyze and think through commands.
\textbf{(2) Auto-Thinking Agent} (e.g., AutoGPT~\cite{AutoGPT[32]}, DemoGPT\footnote{https://github.com/melih-unsal/DemoGPT}) automatically generates code after analyzing/decomposing tasks based on user requirements. However, their relatively closed autonomy prevents users from debugging during operations, making these agents susceptible to internal process freezing.
\textbf{(3) Multi-Agent Collaboration Agent} (e.g., MetaGPT~\cite{DBLP:journals/corr/abs-2308-00352[2]2,hong2024data}, ChatDev~\cite{DBLP:journals/corr/abs-2307-07924[1]2,qian2023experiential,qian2024iterative,qian2024scaling}): Different agents play different roles and collaborate to fulfill user requirements. These agents contain complex intermediate processing steps and prioritize documentation and procedures, failing to maintain user control.
\textbf{(4) Questioning Agent} (e.g., GPT-Engineer~\cite{Gpt-engineer[4]2}, GPT-pilot\footnote{https://github.com/Pythagora-io/gpt-pilot}): These agents generate questions related to user requirements and ask users, expanding the requirements and automatically generating the software code. Questioning Agent attempt to unearth more detailed requirement descriptions from users, but answering these programming-oriented questions may require users to have substantial domain knowledge.

These methods mentioned above involve various ways of completing requirements. However, they lack acceptance criteria during the completion of user requirements. For example, a user requirement stating, "As a traveler, I want to be able to cancel all orders (including flights, hotels, and tickets) in one operation" may appear to an AI agent as a simple database deletion task. But, the real business logic might also include acceptance criteria like "prompting the user for confirmation before canceling the order, needing to cancel 24 hours in advance, etc." The absence of such acceptance criteria in requirements could lead to agents generating software that fail to meet the implicit needs of end-users. 
Moreover, these agents follow the waterfall model~\cite{DBLP:journals/corr/abs-2307-07924[1]2}, characterized by a top-down, sequentially linked order, which easily allows the propagation of biases from earlier to later stages. Especially for software development agents based on large language models, the inevitable hallucination issues of large language models can spread and accumulate within the waterfall model, leading to the generation of code that does not align with user requirements.

The Agile methodologies has the advantages of user collaboration and lightweight iterative feedback~\cite{10.1145/3617169}. Iteration can reduce the accumulation of illusionary errors in large language models that are common in the waterfall model. Incorporating user collaboration into automated software development is not straightforward and requires careful planning of the optimal timing and scenarios for user involvement, as well as the design of participation mechanisms that promote effective interaction. End-users with business needs are constrained by domain knowledge and often find it difficult to describe requirements from an agent development perspective. Thus, the challenge in introducing user collaboration stems from the gap between the semantics of user requirements and the code semantics generated by agents. As shown in Figure~\ref{fig:fig0}, we aim to enable end-users and agents to handle their areas of good and finish software development collaboratively. To achieve this goal, we have built a bridge based on the Agile methodologies for collaborative software development between users and agents. Assuming users have certain expectations for the software they desire when proposing requirements. Our idea design end-users are positioned at both ends of the entire generation process (specifically, requirement clarification and iterative acceptance), with agent participation in the middle. This is because end-users are more adept at decision-making and acceptance suggestions, while agents excel at executing auto-generated tasks.

In this paper, we introduce empowering agile-based generative software development through human-AI collaboration, named AgileGen. First, to address the issue of failing to meet users' implicit needs due to the lack of acceptance criteria, our method complements acceptance criteria by the Gherkin language from the incomplete requirements of users that, is a formal Behavior Driven Development (BDD) language.
Second, to introduce user collaboration and reduce users' learning costs and knowledge constraints, we convert requirements with Gherkin into natural language scenarios through an interaction bridge for requirement clarification. AgileGen has designed three key decision-making processes: requirement proposal, clarification, and iterative acceptance with recommendations, focusing on the skills where end users excel.
Finally, to ensure the consistency of the generated code with user requirements, AgileGen converts the scenarios of user decisions back into Gherkin, participates in the subsequent stages of visual design, and generates consistency factors. The consistency factors guide code generation to ensure the software meets user needs. Furthermore, AgileGen employs a lightweight iterative approach to rapidly present software prototypes to end-users for acceptance or recommendation decisions, facilitating the next iteration. This process leads to a consistent alignment with user requirements through iteration.

Considering the convenience of evaluating the software application for participants, we used web development to demonstrate our idea of the framework design due to the broad compatibility of web-based programming languages. It is worth noting that the design principles of our framework are not limited to web application development. We evaluated AgileGen on 40 web projects and the ``SRDD" software task dataset with diverse requirements. Additionally, through the evaluation of participants, it achieved high user satisfaction scores in the User Experience Questionnaire (UEQ) and Likert scale assessments. Our contributions are as follows:

\begin{itemize}
    \item This paper proposes the AgileGen framework, marking the first integration of agile methodologies with human-AI teamwork for generative software development. It tackles the challenge of addressing users' implicit needs by completing acceptance criteria with the Gherkin language, bridging the gap between incomplete user requirements and precise software functionalities.
    \item AgileGen significantly advances user collaboration by translating Gherkin-based requirements into natural language scenarios through an innovative interaction bridge and decision-making method. This method effectively reduces users' knowledge barriers, emphasizing their strength in decision-making processes such as requirement proposals, clarification, and iterative acceptance. Thus, it addresses the critical challenge of user involvement in software development. 
    \item We introduce a novel approach to maintaining consistency between generated code and user requirements. AgileGen ensures that the generated software aligns with user needs by utilizing consistency factors. This approach reduces discrepancies between user requirements and the generated software, facilitated by designing a lightweight iterative process for rapid prototyping and user feedback.  
    \item We introduce a novel memory pool of decision-making results that provides a more reliable and efficient way for users to refine their requirements. This memory pool shown an approach to incorporating user feedback and experiences in the development process.  
    \item The experimental results have validated the efficiency of AgileGen in generative software development on 40 diverse projects and ``SRDD" software projects. Furthermore, we have developed AgileGen as a user-friendly interactive system, and its source code is available\footnote{https://github.com/HarrisClover/AgileGen}. 
\end{itemize}

\section{MOTIVATION \& Innovation}

\subsection{Motivation}
Implementing localized intelligent assistance using large language models in the software development process is relatively easy. However, users guiding large language models through conversational interaction to generate complete software code require the ability to express and decompose requirements clearly. This process is time-consuming and may demand that users have a comprehensive grasp of the software design and implementation details~\cite{10.1145/3643674}. This limitation makes large language models more commonly used as copilots for developers.

Increasingly, researchers are focusing on extending the development capabilities of large language models to user-defined software development~\cite{10.1145/3641399.3641403}. This generative software development inputs user requirements and produces softwares as output. User requirements are abstract statements written in natural language with accompanying informal diagrams. They specify what services (user functionality) the system is expected to provide and any constraints. Collected user requirements often appear as a ``concept of operations" (Conops) document. In many situations, user stories can play the role of user requirements~\cite{laplante2022requirements[5]2}. For example: "The system should accurately calculate total sales, including discounts, taxes, refunds, and rebates," or "The system should randomly select names from a list of students."

\begin{figure}[t]
	\centering
	\subfigure[Requirement specification notation prevalence~\cite{DBLP:journals/software/KassabL22[6]2}.]{%
		\centering
		\includegraphics[width=0.45\textwidth]{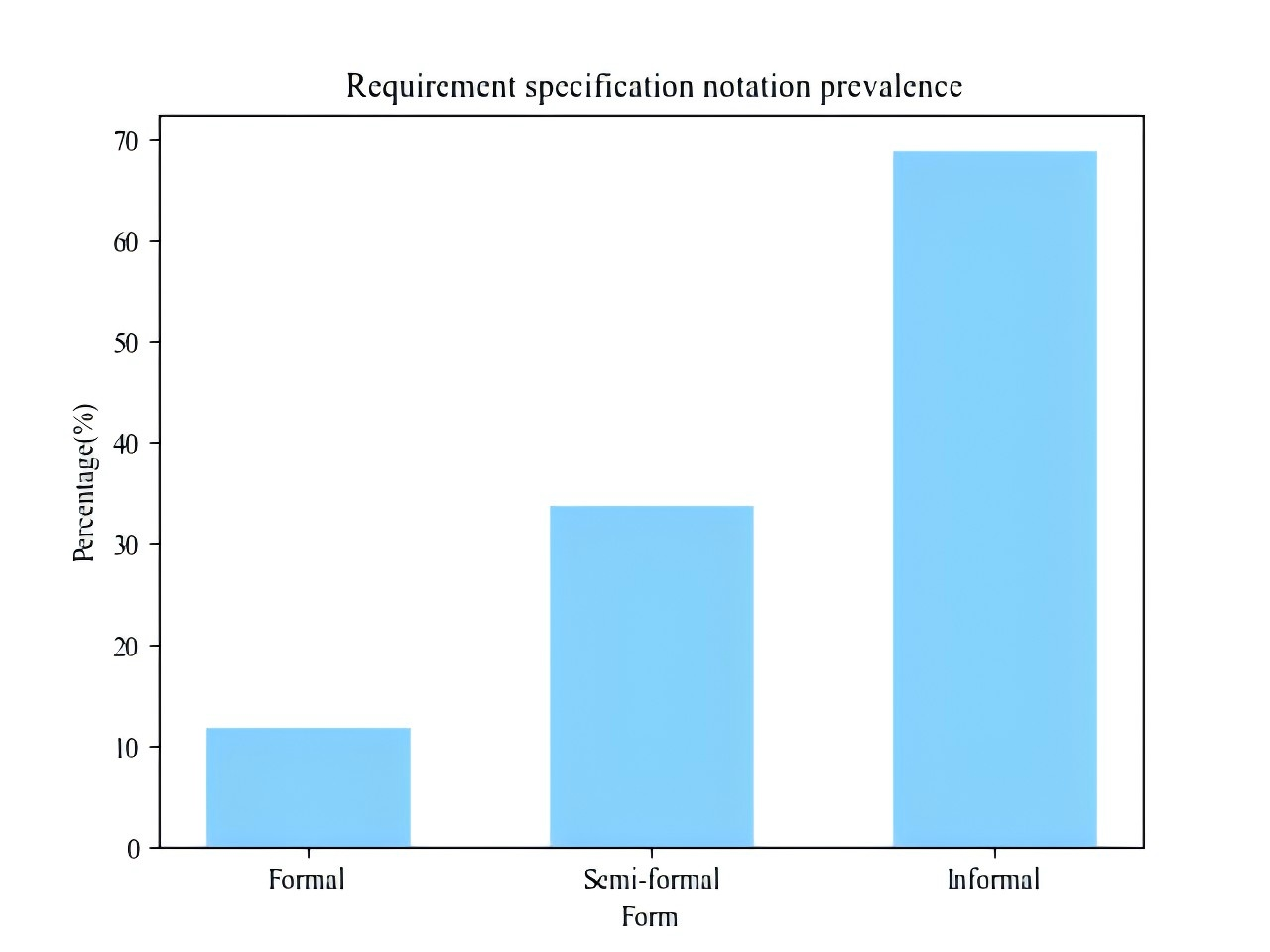}
		\label{fig:sec_2_fig1_a}
	}
	\subfigure[ Level of influence of the human aspects on
the performance of individuals in RE-related activities:
Practitioners’ perspective~\cite{laplante2022requirements[5]2}.]{%
		\centering
		\includegraphics[width=0.45\textwidth]{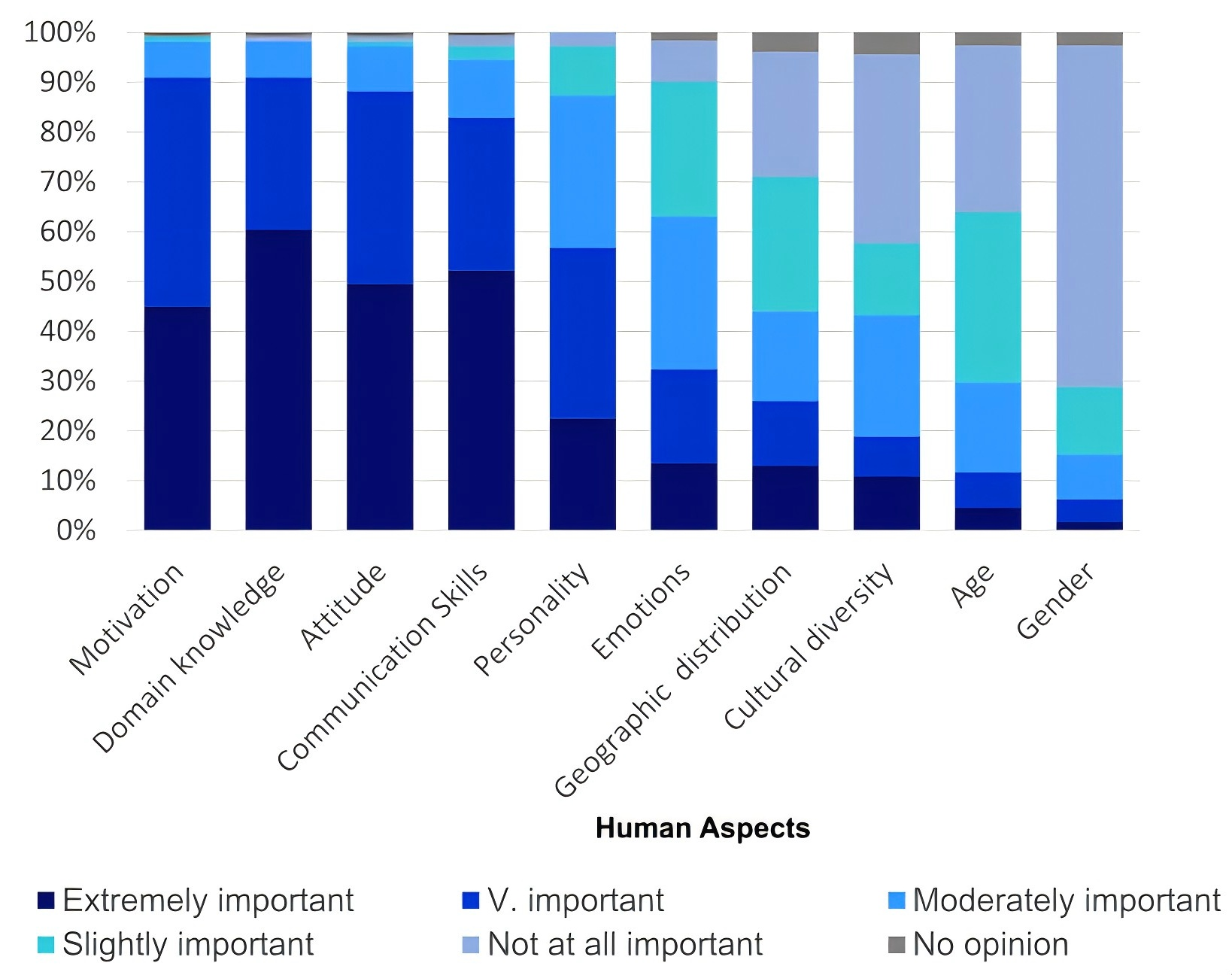}
		\label{fig:sec_2_fig1_b}
	}%
	\centering
        \label{fig:sec_2_fig1}
	\caption{Requirement form \& Human aspects on the performance of individuals in Requirements Engineering (RE)-related activities}
\end{figure}

According to a survey on the popularity of requirement specification symbols conducted by Kassab and Laplante in 2022~\cite{DBLP:journals/software/KassabL22[6]2}, as shown in Figure~\ref{fig:sec_2_fig1_a}, 69\% of respondents in requirements engineering surveys indicated that requirements are expressed in natural language, i.e., informally. Therefore, problems exist in the raw requirements collected from users, including incomplete, as noted by Laplante in 2022~\cite{laplante2022requirements[5]2}. The issues in these raw requirements can lead to the developed software failing to meet the user's acceptance criteria.
\begin{figure}[t]
	\centering
	\subfigure[Communication between end-users and developers is often facilitated through a product manager.]{%
		\centering
		\includegraphics[width=0.45\textwidth]{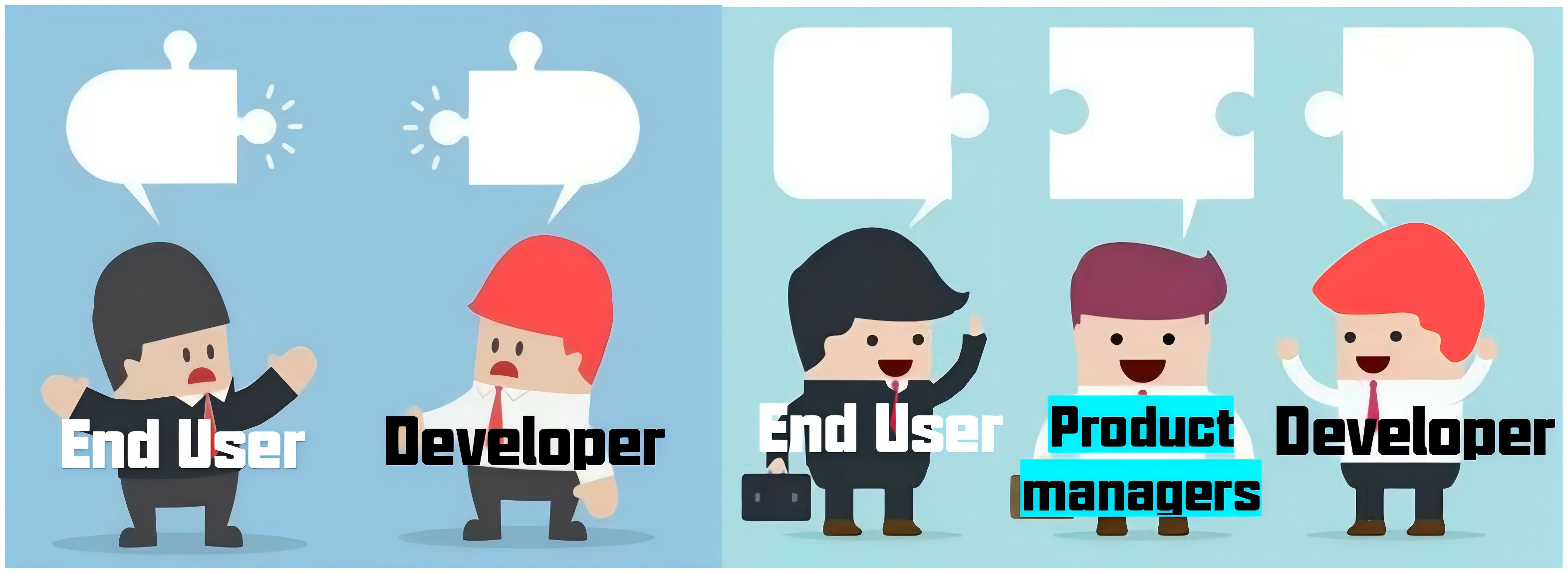}
		\label{fig:sec_2_fig2_a}
	}
	\subfigure[When incomplete requirements from users are directly inputted into AI, the software generated can deviate from the expected.]{%
		\centering
		\includegraphics[width=0.45\textwidth]{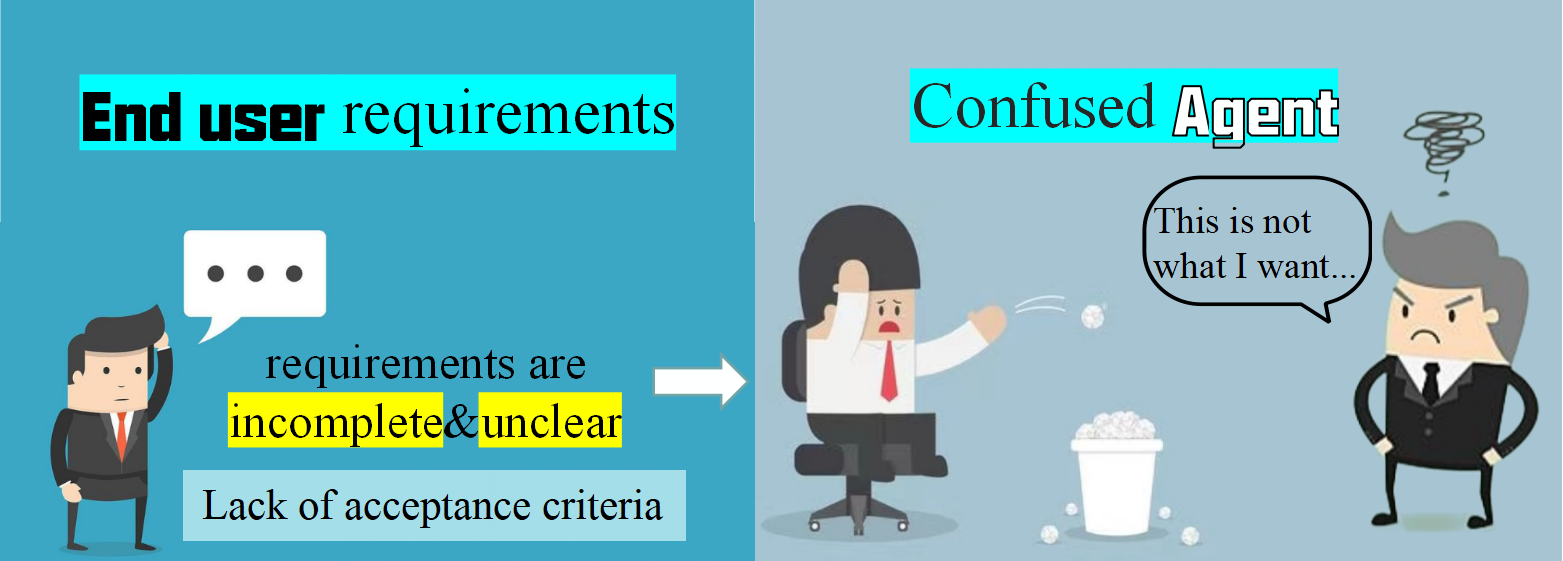}
		\label{fig:sec_2_fig2_b}
	}%
	\centering
        \label{fig:sec_2_fig2}
	\caption{Comparing a software company with an AI-based agent in software development.}
\end{figure}

Typically, software companies overcome this problem by hiring product managers or UI/UX designers to communicate with end users repeatedly and develop detailed requirements analysis documents, as shown in Figure~\ref{fig:sec_2_fig2_a}. Then, developers write code based on these professional requirements documents, and the code undergoes a series of tests before it is maintained or delivered to the customer for acceptance. 

In generative software development, no product manager can assist in analyzing user requirements. Users interact directly with intelligent agents. According to a 2023 study by Dulaji and John~\cite{DBLP:journals/tosem/HidellaarachchiGHM23[7]2} on the human aspects of requirements engineering, ``domain knowledge" is considered extremely important by 60\% of participants and important by 31\% (as shown in Figure \ref{fig:sec_2_fig1_b}). However, end-users often lack domain knowledge and struggle to articulate their requirements clearly. This can lead to discrepancies between the functionalities of the generated software and the end user's expectations (as indicated in Figure \ref{fig:sec_2_fig2_b}).

\begin{table}[t]
\scriptsize
\caption{A comparison of capabilities across ChatGPT, Copilot Chat, Code Interpreter, AutoGPT, DemoGPT, MetaGPT, ChatDev, GPT-Engineer, and gpt-pilot. Please note that "\ding{52}" indicates a specific feature within each framework.}
\begin{tabular}{lcccccccc}
\hline
                & Passive Agent                                                                                                  & \multicolumn{2}{c}{Auto-Thinking}                                                                           & \multicolumn{2}{c}{Multi-Agent}                                                                            & \multicolumn{2}{c}{Questioning Agent}                                                                             & \multicolumn{1}{l}{Huam-AI Teamwork}                      \\ \hline
\textbf{Framework Capabilities}   & \multicolumn{1}{c|}{\begin{tabular}[c]{@{}c@{}}ChatGPT\\ Copilot Chat\\ Code Interpreter\end{tabular}} & AutoGPT                                           & \multicolumn{1}{c|}{DemoGPT}                      & MetaGPT                                           & \multicolumn{1}{c|}{ChatDev}                      & GPT-Engineer                                       & \multicolumn{1}{c|}{gpt-pilot}                    & \multicolumn{1}{c}{AgileGen(Our)}            \\ \hline
User Story Generation          &                                                                                                        &                                                   &                                                   & {\color[HTML]{111111} \ding{52}} &                                                   & {\color[HTML]{111111}  \ding{52}} & {\color[HTML]{111111}  \ding{52}} & {\color[HTML]{111111}  \ding{52}} \\
Gherkin Scenario Generation     &                                                                                                        &                                                   &                                                   &                                                   &                                                   &                                                   & {\color[HTML]{111111} }                           & {\color[HTML]{111111}  \ding{52}} \\
Scenario Decision (Confirm)          &                                                                                                        &                                                   &                                                   &                                                   &                                                   &                                                   & {\color[HTML]{111111}  \ding{52}} & {\color[HTML]{111111}  \ding{52}} \\
Scenario Decision (Add)          &                                                                                                        &                                                   &                                                   &                                                   &                                                   &                                                   & {\color[HTML]{111111}  \ding{52}} & {\color[HTML]{111111}  \ding{52}} \\
Scenario Decision (Delete)          &                                                                                                        &                                                   &                                                   &                                                   &                                                   &                                                   &                                                   & {\color[HTML]{111111}  \ding{52}} \\
Scenario Decision (Modify)          &                                                                                                        &                                                   &                                                   &                                                   &                                                   &                                                   &                                                   & {\color[HTML]{111111}  \ding{52}} \\
Visual Design         &                                                                                                        &                                                   &                                                   & {\color[HTML]{111111}  \ding{52}} &       {\color[HTML]{111111}  \ding{52}}                                            &                                                   &                                                   & {\color[HTML]{111111}  \ding{52}} \\
Consistency Factor        &                                                                                                        &                                                   &                                                   &                                                   &                                                   &                                                   &                                                   & {\color[HTML]{111111}  \ding{52}} \\
Code Generation            & {\color[HTML]{111111}  \ding{52}}                                                      & {\color[HTML]{111111}  \ding{52}} & {\color[HTML]{111111}  \ding{52}} & {\color[HTML]{111111}  \ding{52}} & {\color[HTML]{111111}  \ding{52}} & {\color[HTML]{111111}  \ding{52}} & {\color[HTML]{111111}  \ding{52}} & {\color[HTML]{111111}  \ding{52}} \\
Code Execution            &                                                                                                        &                                                   &                                                   &                                                   &                                                  {\color[HTML]{111111}  \ding{52}} & {\color[HTML]{111111}  \ding{52}} & {\color[HTML]{111111}  \ding{52}} & {\color[HTML]{111111}  \ding{52}} \\
Acceptance \& \\ Recommendation        &                                                                                                        &                                                   &                                                   &                                                   &                                                   &                                                   &                                                   & {\color[HTML]{111111}  \ding{52}} \\
Interactive Interface            &                                                                                                        &                                                   & {\color[HTML]{111111}  \ding{52}} & {\color[HTML]{111111}  \ding{52}} &  {\color[HTML]{111111}  \ding{52}} &                                                   &                                                   & {\color[HTML]{111111}  \ding{52}} \\ \hline
\textbf{Input-Output Format} &                                                                                                        &                                                   &                                                   &                                                   &                                                   &                                                   &                                                   &                                                   \\
Input Requirement \\(One-line/Detailed)     & Detailed                                                                                                     & Detailed                                                & Detailed                                                & One-line                                              & One-line                                                & Detailed                                                & Detailed                                                & One-line                                                \\
Output (Code Snippet/Project)      & Code Snippet                                                                                                    & Project                                                & Project                                                & Project                                                & Project                                                & Project                                                & Project                                                & Project                                                \\ \hline
\end{tabular}
\label{table:related_agents_func}
\end{table}

In generative software development, intelligent agents with GitHub stars >= 1k are categorized into four types: passive agent, auto-thinking agent, multi-agent collaborative agent, and questioning agent. The goal is to assist developers in rapidly developing code prototypes. The functionalities of each type of intelligent agent are compared in Table \ref{table:related_agents_func}. Auto-thinking and multi-agent collaborative agents adopt the waterfall model for development, generating software code end-to-end (e.g., in AutoGPT, user participation is limited to authorization actions). These types of agents can automatically generate project code based on user descriptions. However, agents entirely reliant on the pre-training capabilities of large language models tend to accumulate errors resulting from illusions. These problems become more pronounced in the face of incomplete user requirements, potentially leading to deviations from the users' acceptance criteria.

\begin{figure}[t]
	\centering
	\includegraphics[width=1\textwidth]{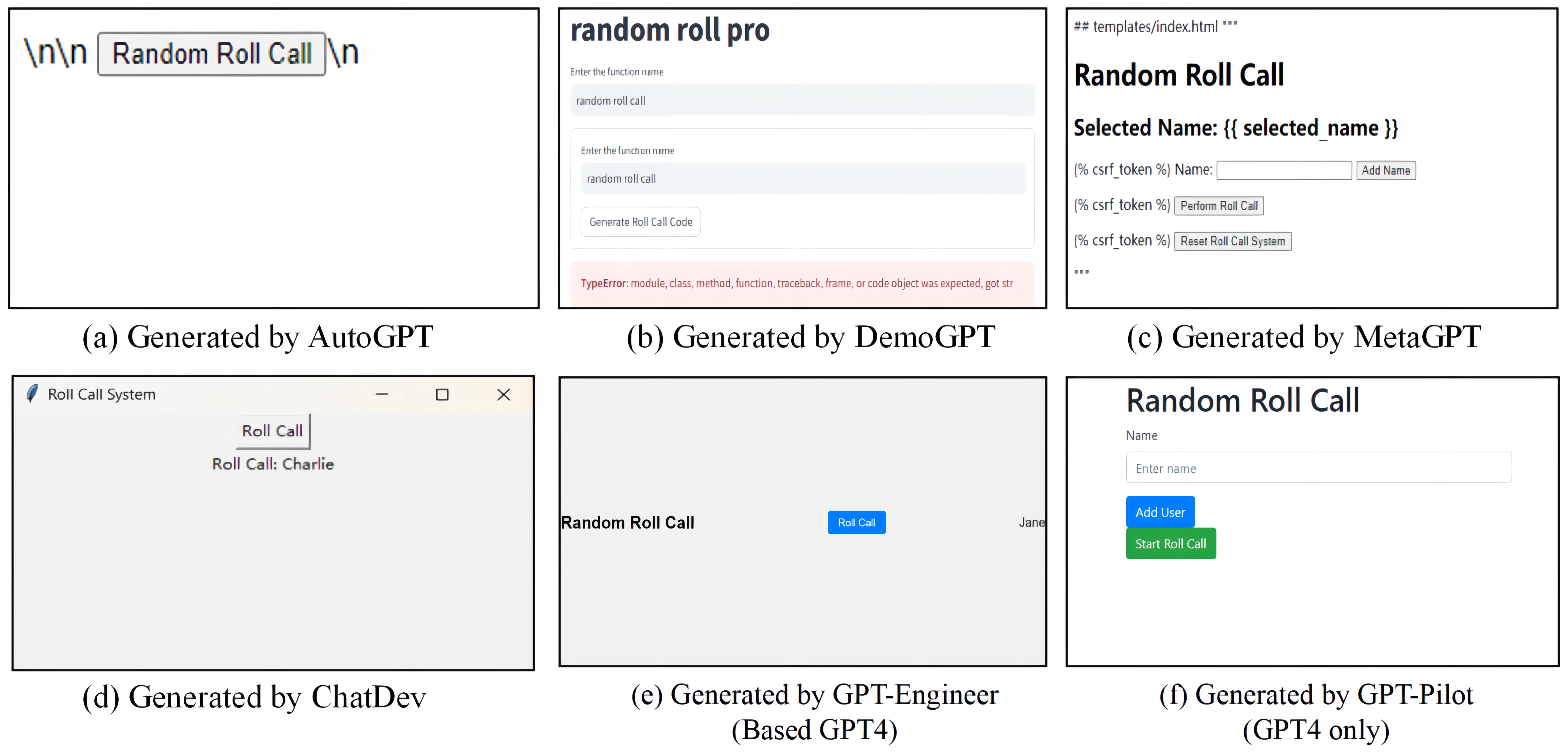}
	\caption{\label{fig:sec_2_fig4} Display of the code generation effects by AutoGPT, DemoGPT, MetaGPT, and ChatDev agents for the requirement ``Please generate a web system code with random roll call function".}
\end{figure}

Taking the practical classroom requirement "Please generate a web system code with random roll call function." as an example, the function is to select a student name from a list to answer questions randomly. As shown in Figure \ref{fig:sec_2_fig4}, different agents input the same requirement description and use GPT3.5/GPT4 to generate code ten times, selecting the best-performing example for execution. Among these, the code generated by AutoGPT and DemoGPT requires manual adjustments to be executable. As shown in Figure~\ref{fig:sec_2_fig4}(a) the interface generated by AutoGPT is rudimentary and cannot perform functions properly. DemoGPT, relying on the ``Streamlit" visualization tool, produces a more aesthetically pleasing interface. However, as shown in Figure \ref{fig:sec_2_fig4}(b), it deviates from the user's requirement. The user expects the function of a random roll call, not an AI used to generate a roll call code. 
As shown in Figure~\ref{fig:sec_2_fig4}(c), the interface generated by MetaGPT is also quite rudimentary. Although it includes an "Add Name" feature, it still cannot complete the roll call function normally. ChatDev generates code using the Python GUI tool Tkinter. It can perform the basic random roll call function as shown in Figure \ref{fig:sec_2_fig4}(d), but the list of names for the random roll call is not operable.

Questioning agent, including GPT-Engineer and GPT Pilot, operates by posing questions to users and waiting for their input. After the user responds to all questions, the agent automatically generates software code based on all inputs. Each user story generated by GPT Pilot is subject to user decisions for confirmation or augmentation, as shown in Table \ref{table:related_agents_func}. Questioning agent expands raw requirements by asking users questions, but some queries may leave users uncertain about how to respond. For example, "How should the database be structured?" or "What should be the criteria for defining a random roll call?". Figures \ref{fig:sec_2_fig4}(e) and \ref{fig:sec_2_fig4}(f) illustrate their performance on random roll call tasks. GPT-Engineer, based on GPT4, generated HTML-based code, similar to ChatDev, achieving basic random functionality but with incomplete name list features. GPT Pilot, unable to implement code on GPT-3.5, was only tested on GPT-4\footnote{https://github.com/Pythagora-io/gpt-pilot/issues/187}. It showed the added functionality for the name list, but the button click had no response\footnote{The generation process is prone to freezing, especially after completing multiple subtasks https://github.com/Pythagora-io/gpt-pilot/issues/38}.

Agents based on the top-down workflow of the waterfall model are prone to accumulating bias or process freezes, with no room for human intervention in the intermediate stages. Questioning agents, by asking users questions, can explore implicit requirements when users know how to respond. However, not all questions are reasonable, and this way to expand user requirements challenges the users' domain knowledge. As generative software development becomes more mainstream, the agents for end users with different domain knowledge need to be more user-friendly. As indicated by the statistics from the generative software development agents GPT-pilot GitHub Issues 655\footnote{https://github.com/AntonOsika/gpt-engineer/issues/655} and 68\footnote{https://github.com/Pythagora-io/gpt-pilot/issues/68}, end-users are more adept at scenario decision-making and iterative acceptance and recommendation, while Large Language Models are better at implementing code.

In summary, we advocate for a human-AI collaborative generative software development agent oriented from the end-user's perspective. This agent would enable end users and agents to handle their respective strengths, bridging the gap between end-user requirements and the technical implementation by agents. The objectives to be achieved are: (1) To reduce the involvement of user domain knowledge while being able to explore the implicit requirements of end-users. (2) To ensure the consistency of the generated software with user requirements. (3) To present a user-friendly interaction interface for end users.

\subsection{Innovation}

Our innovations are divided into three folds:
First, at the methodological level, we propose a human-agent collaborative generative software development approach, focusing on human involvement at the beginning (scenario decision-making) and end (acceptance \& recommendation decision-making) of each iteration, with the agent contribution to the intermediate steps, as illustrated in Figure \ref{fig:fig0}. Unlike fully automated methods like AutoGPT, DemoGPT, MetaGPT, and ChatDev, which can accumulate errors and biases from large language models, our approach maintains user control and minimizes errors. Methods like GPT-Engineer and gpt-pilot require extensive human input for requirements clarification but don't ensure complete business logic coverage and demand high technical knowledge. Our method leverages the strengths of both humans and agents, optimizing collaboration to complete software development tasks efficiently while reducing human involvement costs. For more on the human-agent collaboration experience in our method, see section RQ3 \ref{Sec:RQ3}.

Second, at the software design level, we are the first to introduce the core concepts and processes of behavior-driven development (BDD) into generative agents to accomplish generative software development tasks. Using the Gherkin language of BDD to generate user scenarios and complete acceptance criteria within user requirements, we bridge the gap between incomplete user requirements and precise software functionalities. The acceptance \& recommendation decision-making of end-users is designed to reflect the agile development philosophy of incremental delivery of small versions, dynamically incorporating human decision-making into version iterations. Other methods typically adopt an end-to-end model approach, where errors in earlier steps can easily propagate to subsequent steps, resulting in software not aligning with user requirements.

Lastly, in our approach design and implementation, we extend the concept of AI-Chain~\cite{10.1145/3638247} to agent graph composed of carefully designed prompts, forming a directed cyclic graph connected by well-designed prompts, as shown in Figures~\ref{fig:scDesign} and~\ref{fig:scRapid} in section~\ref{secpattern}. This agent graph includes human decision-making points to facilitate iterative processes within the cycles. The design of this agent graph is highly scalable. Adding new functionalities only requires only consideration of the preceding component's output and the next one's input, enabling seamless integration. If human decision points are bypassed, the method becomes fully end-to-end generative. We introduce a memory pool to collect human decision results and recommend them to future tasks with similar requirements, creating a self-updating loop. To reduce the learning cost of human interaction, we designed an interaction bridge between the Gherkin language and natural language.
Additionally, we use Gherkin scenarios as consistency factors to guide code generation. Compared to other methods, as shown in Table~\ref{table:related_agents_func}, our approach includes eight successful visual design principles to guide visual design generation, including human decision-making like scenario addition, deletion, modification, and acceptance \& recommendation. While other methods, such as MetaGPT, GPT-Engineer, and gpt-pilot, generate user stories by posing questions based on predefined rules~\footnote{https://github.com/Pythagora-io/gpt-pilot/blob/main/core/prompts/spec-writer/ask\_questions.prompt}, we generate user stories using Gherkin syntax.

\section{Design of Human-AI Collaborative AgileGen}

In this section, we first outline the architectural design of AgileGen, an Agile-based generative software development through human-AI teamwork. Then, in Section \ref{sec:decision-making}, we will introduce the end-user decision-making and interaction interface components in AgileGen. Finally, in Section \ref{secpattern}, we will detail the design of the core components of AgileGen. We use website application development as an example to demonstrate how AgileGen navigates end-user decisions through various components to generate a website application that aligns with user requirements.
Additionally, this section provides an example to illustrate how AgileGen bridges collaborative development between end users and large language models, aiming to fulfill a single-line requirement: "Please generate a web system with random roll call function."

\subsection{AgileGen Overview}
\begin{figure*}[h]
\centerline{\includegraphics[width=\textwidth]{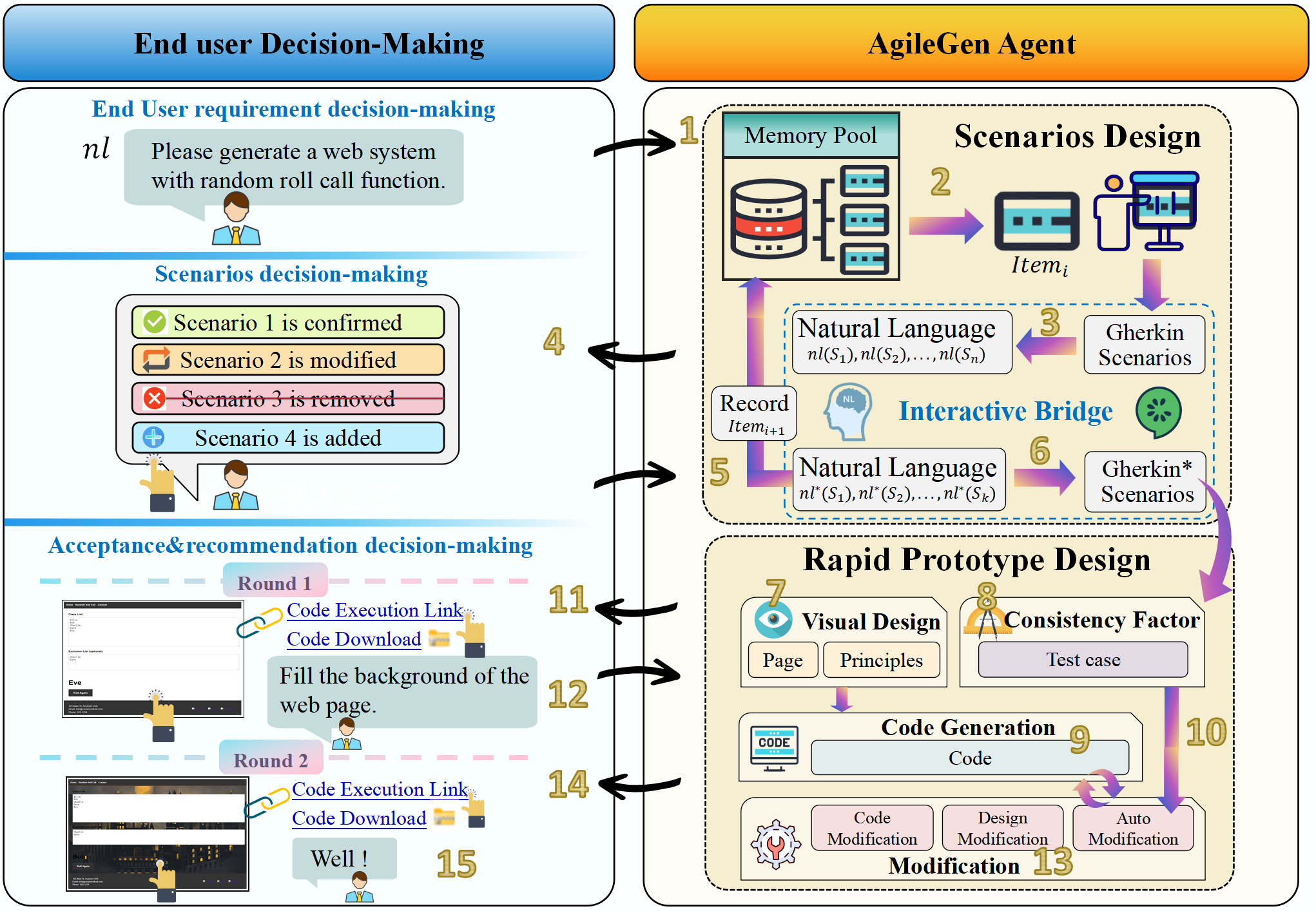}}
\caption{AgileGen Overview Diagram. The left side displays the user decision-making part, while the right shows the AgileGen Agent's automatic generation. The black arrow in the middle represents the interaction between the user and the Agent. The process starts with the End User requirement decision, followed sequentially by the Scenario Design component, Scenario decision-making, Rapid Prototype Design, and Acceptance \& recommendation decision-making. The outcome is the software code.}
\label{fig:agile sapper}
\end{figure*}

As shown in Figure \ref{fig:agile sapper}, the design of AgileGen is divided into two parts. The ``End-user Decision-Making" component is dedicated to collaborating with end-users to collect and clarify their decisions. The AgileGen Agent responds to user decisions and transforms and analyzes them to guide large language models to generate software code that aligns with user requirements. 
``Scenarios Design" and ``Rapid Prototype Design" are two core components of ``AgileGen". The Scenarios Design component is primarily used for designing different scenarios represented in the Gherkin language based on requirement decision-making. These scenarios are presented to end-users for scenario decision-making, returning the decided Gherkin scenarios. The Rapid Prototype Design component mainly focuses on generating software code guided by the decided Gherkin scenarios, presenting the software to users for acceptance, receiving user recommendations, and modifying the code accordingly. Gherkin is a formal Behavior Driven Development (BDD) language that uses keywords to structure and define executable specifications for the user story~\cite{Parsa_Saeed[58]}. Gherkin focuses on features that deliver business value and helps define acceptance criteria for end-user requirements. Scenarios expressed in Gherkin language supplement the acceptance criteria for user requirements. Still, the formalized definition also clearly specifies the content of code generation, facilitating the creation of code that aligns with user requirements.

Next, we will simulate the AgileGen agent's process. Initially, the end-user inputs a requirement described in a single line of natural language ($nl$). This requirement is received by the Scenarios Design component, with the workflow as follows:

(1) Match task entries similar to $nl$ in the ``Memory Pool".

(2) Extract the entry with the highest semantic similarity that exceeds the threshold from the memory pool to use as a reference for designing Gherkin scenarios.

(3) Convert the Gherkin scenarios, represented in a domain-specific language, into natural language scenarios.

(4) Present the natural language scenarios to the user in a list form, allowing the end-user to make decisions in the scenario list (confirm/add/delete/modify).

(5) The decided natural language scenarios, as domain knowledge specific to the requirement $nl$, are recorded in the memory pool along with $nl$.

(6) The decided natural language scenarios are converted back into Gherkin language ($\mathrm{ Gherkin^* }$) and received by the next component.

Then, the decided $\mathrm{ Gherkin^* }$ is received by the Rapid Prototype Design component, with the workflow as follows:

(7) Plan visual design guided by $\mathrm{ Gherkin^* }$.
    
(8) Generate consistency factors consistent with the requirements guided by $\mathrm{ Gherkin^* }$.
    
(9) Integrate visual design and $\mathrm{ Gherkin^* }$ to guide large language models in generating code.
    
(10) Input the generated code and the consistency factor into the code modification module to automatically perform code modification once.
    
(11) The generated code interacts with the end-user through the ``Code Execution Link" for execution.
    
(12) The end-user accepts the software or inputs feedback recommendations for improvements.
    
(13) Accept the recommendations and modify the code.

(14) The end-user accepts the software or inputs feedback for improvements. 
    
(15) This continues until the end-user is satisfied and downloads the compressed code files via ``Code Download".

The design of AgileGen positions the involvement of the end user at both ends of the process. As shown in Figure~\ref{fig:agile sapper}, the ``End user Decision-Making" of requirement clarification (Requirement decision-making and Scenarios decision-making) and iterative acceptance (Acceptance\&recommendation decision-making). The automatic code generation by agents is situated in the intermediate phase. This design allows end-users to easily and conveniently clarify requirements lacking acceptance criteria with AgileGen. The scenario design represented by Gherkin fills in the missing acceptance criteria, and the consistency factors generated from these criteria positively influence the automation of code generation. Finally, an executable software prototype is presented to the end-user, who, after providing acceptance recommendations, enters the next round of rapid prototype development.

\subsection{User Decision-Making \& Interaction Interface Design}
\label{sec:decision-making}

In this section we first describe how end-users decision-making can be imported into the AgileGen architecture and used to enhance the whole generative architecture continuously. Then, we describe the design of the interaction interface presented to the end-user. 
\subsubsection{User Decision-Making}
\label{enduser}
As in Figure~\ref{fig:agile sapper}, in the AgileGen architecture, the end-users are involved in (1) End-user requirement decision-making, (2) Scenarios decision-making, (3) Acceptance \& recommendation decision-making.

\textbf{
End-user Requirement Decision-making}.
The natural language requirement of the end-users is the first act of user participation in decision-making that determines the goals of software to be generated.
Specifically, the end-user proposes the natural language requirement ($nl$). The end-user needs to describe at least one sentence or phrase based on the software they want to achieve. Phrases like: "I want to...", "Please help me to generate...", "I would like to..." etc. For example: ``Please generate a web system with a random roll call function".

\textbf{Scenario Decision-making}.\label{sec:Scenario Decision Making}
Scenario decision-making is to allow end-users to make (Confirmed/ Addition/ Deletion/ Modification) decisions on the generated scenarios, which are scenarios of Gherkin generated based on $nl$ and experience of predecessors in memory pool. This process is used for iterative clarification of requirements by end-users.
Specifically:
\begin{itemize}
    \item First, the requirement ($nl$) is matched with the requirement descriptions ($pl_i$) of items stored in the memory pool and selects items if they are semantically similar. Form of each item follow:
\begin{equation}
    Item_i = <pl_i,nl^{'}(S_1),nl^{'}(S_2)\\,...,nl^{'}(S_k)>
\end{equation}
where $nl^{'}(S_k)$ represents the description of the scenario that has been decision-making by predecessors, $S_k$ represents scenario.
$k$ represents the number of scenarios. $i$ represents the entry index, $1\leq i \leq n$, where $n$ represents the total number of entries in the Memory Pool.
    \item Second, the selected $Item_i$ serves as a reference in the design of Gherkin scenarios. In practice, presenting the Gherkin language directly to the end-user may result in a suboptimal experience, as they may lack the requisite domain-specific language knowledge. Therefore, we designed the Interaction Bridge to create a bi-directional transition between formal Gherkin syntax and natural language scenario descriptions.
    \item Finally, the split scenarios $S_1,S_2,\ldots,S_n$  from the Gherkin language description are translated to natural language scenarios by the interaction bridge: $nl(S_1),\textit{nl}(S_2)\textit{...} ,\textit{nl}(S_n)$. The end-user makes decisions on these scenarios, and these scenarios denoted by  $nl^{*}(S_1),nl^{*}(S_2),... ,nl^{*}(S_k)$ are recorded as a new human experience knowledge item in the form of Eq~\ref{eq:2} within the memory pool.
    \begin{equation}
    \label{eq:2}
        Item_{i+1} = <nl,nl^{*}(S_1),nl^{*}(S_2),...,nl^{*}(S_k)>
    \end{equation}
\end{itemize}

The experience database stores many scenario descriptions for specific requirements through this process. Furthermore, these human experiences accumulate with increased usage. The human experience guides the agent during software applications generation, enhancing the system to produce more integrally and reliably user stories.

\textbf{Acceptance and Recommendation Decision-making.}

During the acceptance \& recommendation, our AgileGen offers a ``Code Execution Link". Opening the link reveals the interface of the software. End-users can test the function shown in the software's interface, allowing them to make recommendations and adjustments.
The final interface of the random roll call website is refined through iterative user recommendations, aligning the style with end-user preferences.
The final generated code will be packaged and available to users through a ``Code Download link".

\begin{figure*}[h]
\centerline{\includegraphics[width=\textwidth]{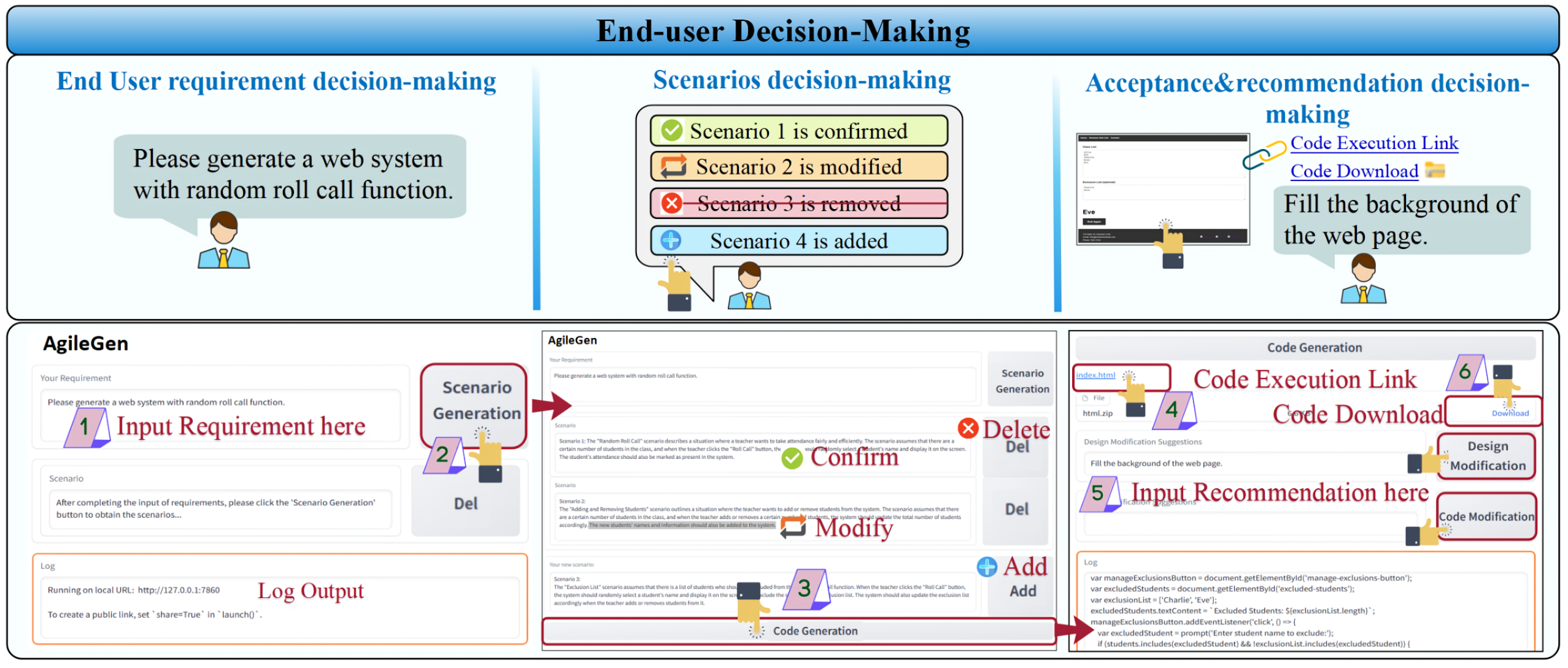}}
\caption{AgileGen Interaction Interface that helps End-User Decision-Making.}
\label{fig:interface}
\end{figure*}

\subsubsection{AgileGen Interaction Interface Design}
\label{sec:interface}

The AgileGen interaction interface is built on the Gradio\footnote{https://www.gradio.app/} open-source library, which offers high extensibility. The overall design principles include clarity, ease of use, and a clear objective. Figure~\ref{fig:interface} demonstrates the process of interaction between the end-user and the AgileGen interface. Next, we will introduce the AgileGen interface design according to three categories of end-user decision-making.

\textbf{End User requirement decision-making.} 
The initial interface of AgileGen is divided into three areas: the Requirement Input Area, the Scenario Generation Area, and the Log Output Area.
\textit{Requirement Input Area}: This is designated for end-users to type in their raw requirements. Following the principle of consistency with user habits, the initial ``Textbox" is placed on the left side, with action buttons on the right. Additionally, it includes a user-friendly prompt when no input has been made yet: ``Please input your requirement here...".
\textit{Scenario Generation Area}: This area displays scenario entries and allows users to make decisions for each entry. In the initial interface, an empty, non-editable textbox is pre-initialized with placeholder content advising users to input their requirements and then click the “Scenario Generation” button to proceed.
\textit{Log Output Area}: This displays the internal process results of the AgileGen system, initially showing the accessed local URL.
During the requirement decision-making process, the interaction between end-users and AgileGen involves two steps: (1) Enter the requirement in the ``Textbox". (2) Clicking the “Scenario Generation” button.

\textbf{Scenarios decision-making.} 
When the end-user clicks the ``Scenario Generation" button, the waiting time is displayed as ``processing | actual time/estimated time(s)" to show progress, and the log area outputs the system's internal processes. Then, the natural language descriptions of the scenarios are displayed item by item in the Scenario Generation Area, using the regex rule \^{}\textbackslash{}s*(?:Feature|Backgr-ound|Scenario(?: Outline)?|Examples)\textbackslash{}b to display Gherkin scenarios by entry. End-users can check each scenario description individually and conveniently modify the description in the ``Textbox" or click the ``Del" button to delete an entry. Upon completing scenario generation, the AgileGen interface reveals hidden areas to the user, namely the ``Your new scenario" area and the ``Code Generation" button.
\textit{Your new scenario Area}: Includes a Textbox and an "Add" button for end-users to add scenario descriptions actively. End-users type a natural language scenario description in the Textbox, and after clicking the Add button, the new scenario is added to the Scenario Generation Area.
End-users can freely confirm, add/delete/modify scenarios with AgileGen during the scenario decision-making. 
\textit{Code Generation Button}: After deciding on the scenarios, the end-user clicks the ``(3) Code Generation" button to confirm that the decided scenarios will be input into the automatic rapid prototype design process.

\textbf{Acceptance and Recommendation Decision-making.} After the rapid prototype design, the AgileGen interface will display four hidden areas, including the \textit{Code Execution Link (index.html) area}, \textit{Code Download area}, \textit{Design Modification area}, and \textit{Function Modification area}.
``Code Execution Link" area is presented as a hyperlink, and the end-user can click this link to accept and review the prototype design results. If the end-user is satisfied, they can click the link in the Code Download Area to download the packaged code files. Different code files are generally extracted by different rules, with HTML files structured as ``index.html:\textbackslash{}n```html(.*)```\textbackslash{}nend index.html". CSS files as "style.css:\textbackslash{}n```css(.*)```\textbackslash{}nend style.css". Javascript files as "script.js:\textbackslash{}n```javascript(.*)```\textbackslash{}nend script.js".
The modification area includes a Design Modification area and a Function Modification area. Recommends for modifications raised by the end-user during acceptance can be typed into the "Textbox" within the modification area, and the user can click the button to modify the code. Here, we differentiate between interface design and functional suggestions because, for website application development, this will involve modifications to code in different languages. The results of the modifications are still presented in the form of links on the AgileGen interface, allowing users to perform iterative acceptance again through the same operations.

Our AgileGen interface adheres to simplicity and ease of use. The Log output area displays real-time outputs from the AgileGen agent system, which aids developers in maintenance and debugging.

\subsection{Design of Core Components}\label{secpattern}
\begin{figure}[h]
\centerline{\includegraphics[width=0.9\textwidth]{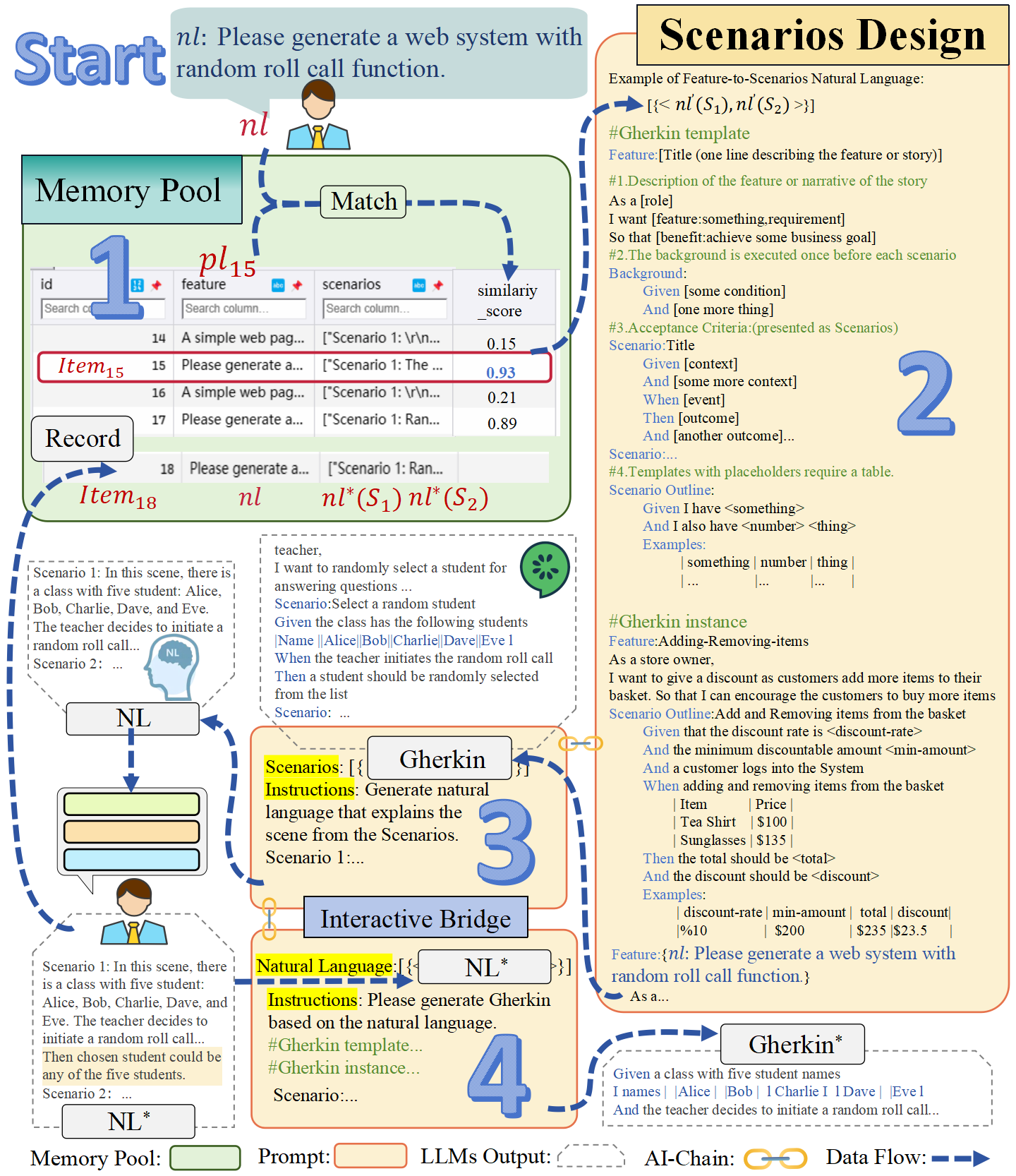}}
\caption{Overview Diagram of the Scenario Design Component. The numbers in the figure are the AgileGen execution sequences. The dotted arrows show the data flow of each component.}
\label{fig:scDesign}
\end{figure}

In this section, we will introduce the core components of AgileGen, including the Scenario Design and Rapid Prototype Design components. Drawing inspiration from the AI-Chain~\cite{10.1145/3638247} concept, we link large language models as components graph capable of performing complex functions. Next, we will describe the design principles and implementation details.

\subsubsection{Scenario Design component}\label{sectionGherkin}

The Scenario Design component is used to supplement the missing acceptance criteria with the end-user to clarify the end-user's requirements. The detailed implementation is shown in Figure~\ref{fig:scDesign}. Here, 1, 2, 3, and 4 represent different functional blocks within the component, with the dashed line enclosing the output of these blocks. After the end-user submits a requirement $nl$, the Scenario Design component follows the sequence of End User requirement decision-making $\rightarrow$ 1: Memory Pool $\rightarrow$ 2: Scenario Design $\rightarrow$ 3: Gherkin to NL $\rightarrow$ End User Scenarios decision-making $\rightarrow$ 4: NL to Gherkin, ultimately outputting the decided $\mathrm{Gherkin^*}$.

\textbf{1: Memory Pool}. The memory pool design is intended to enhance the reliability of scenario design. Scenarios decided by previous users are stored in the memory pool and selected to guide current scenario design in the form of examples. As shown in Figure~\ref{fig:scDesign}(1), when the end-user submits a requirement $nl$, it is compared with the "feature" attribute entries in the memory pool (denoted as $pl_i$,$1\leq i \leq n$) to calculate a similarity score. The matching formula is as follows:

    \begin{equation}
    \label{eq:jacc}
        \operatorname{Jaccard}(t_1, t_2) = \frac{|t_1 \cap t_2|}{|t_1 \cup t_2|}
    \end{equation}
    \begin{equation}
    similariy\_score = \max _{1 \leq i \leq n}\left(\operatorname{Jaccard}\left(n l, p l_i\right)\right)
    \end{equation}
    
The Eq~\ref{eq:jacc} measures the overlap between two texts at the lexical level, where $t_1$ and $t_2$ represent the sets of words obtained from converting two sentences, respectively. The Jaccard similarity calculates the ratio of the intersection to the union of these two sets. The entry with the highest similariy\_score is selected from the memory pool over the threshold. In the figure, $Item_{15}$ is chosen as an example to insert the content of the attribute scenarios into the Scenarios Design.

The memory pool is initially empty and gets populated with data as users make scenario decisions. After a user makes a decision, the user's requirement $nl$ along with the decided scenarios $\mathrm{NL^*} = \{nl^*(S_1), nl^*(S_2)\}$ are recorded in the memory pool as human experiential knowledge. While there may be multiple real scenarios, $S_1$ and $S_2$ are used as examples for convenience. The continuously accumulated human experiential knowledge serves as examples for generating subsequent scenarios. By designing such a memory pool, we aim to enhance the reliability of our system's output through ongoing practice.

\textbf{2: Scenarios Design.}
Agile utilizes user story to acquire demands for software development. These stories require users to specify their role design, functionality vision, and value statement. Acceptance criteria are important for determining whether the user expectations are met before project delivery~\cite{10.5555/2509724[57]}. However, not everyone is good at storytelling. To address this,  Gherkin uses keywords to structure and define executable specifications for the user story~\cite{Parsa_Saeed[58]} and helps define acceptance criteria for end-user requirements. The goal is to generate a well-formed Gherkin language from the $nl$ of the end-user.
The entire scenario design AI-Chain is depicted in Figure~\ref{fig:scDesign}(2), including selected scenario examples from the memory pool, a predefined Gherkin template~\cite{Parsa_Saeed[58]} (Feature:[Title]), a Gherkin instance (Feature: Adding-Removing-items), and the input interface Feature:\{$nl$\}. Moreover, explanatory comments are added to the Gherkin template to guide the large language models in better understanding and generating the user story in Gherkin format corresponding to $nl$, following the Chain-of-thought concept~\cite{DBLP:conf/nips/Wei0SBIXCLZ22[9]2}. 
The generated result includes two scenarios, as shown in the Gherkin dashed expansion box in Figure~\ref{fig:scDesign}. The user story clearly describes the motivation, action, and feedback, with well-defined acceptance criteria.

\textbf{3,4: Interactive Bridge with Gherkin.}
The user story, written in formal Gherkin language, describes the functionalities needed. During the scenarios decision-making, communication with the end-user for clarifications requirements. End-users tend to be more at ease when communicating in non-formal natural language (NL). This section explains how we will design an interaction bridge to eliminate the gap between non-formal natural language and formal Gherkin.

\textbf{3: Gherkin to NL.}
The process of converting the Gherkin language into natural language descriptions is illustrated in Figure~\ref{fig:scDesign}(3). Gherkin from 2: Scenario Design is segmented by rules introduced in Section~\ref{sec:interface}. The segmented scenarios are filled into the input interface of 3: Gherkin to NL as Secnarios:[{$S_{1},S_{2}$}]. This is followed by an instruction we found effective through multiple trials: generating natural language from Gherkin and attempting to explain it. Our agent interpret the ``Scenarios" in this instruction corresponding to the contents after the keyword "Scenarios: ". The generated results are displayed in the NL dashed expansion box in Figure~\ref{fig:scDesign}, with natural language scenarios arranged by sequence number.

\textbf{4: NL to Gherkin.} 
The process of converting the natural language scenarios decided by the user, $NL^*=<nl^*\left(S_1\right), nl^*\left(S_2\right)>$, into Gherkin scenarios is depicted in Figure~\ref{fig:scDesign}(4), consisting of four parts: the natural language input interface, Instruction, Gherkin template, and Gherkin instance. Unlike the ``Scenarios Design", the instruction here focuses on converting natural language into Gherkin scenarios, not the entire feature. The user-decided scenarios $\mathrm{NL^*}$ are input into the ``4: NL to Gherkin" input interface, and recorded in the memory pool along with the end-user's requirement $nl$. For example, as shown in Figure~\ref{fig:scDesign}, the end-user modified Scenario 1 during the scenario decision-making, adding the sentence: "Then the chosen student could be any of the five students." The decided scenario $\mathrm{NL^*}$ and the end-user requirement $nl$ are recorded as $Item_{18}$ and saved in the memory pool. Then, the decided scenarios $\mathrm{NL^*}$ are input into the 4: NL to Gherkin natural language input interface and the final Gherkin scenarios are represented by $\mathrm{Gherkin^*}$. With this, the Scenario Design component completes its task.

\subsubsection{Rapid Prototype Design component.}\label{sectionRapid}
\begin{figure}[t]
\centerline{\includegraphics[width=1\textwidth]{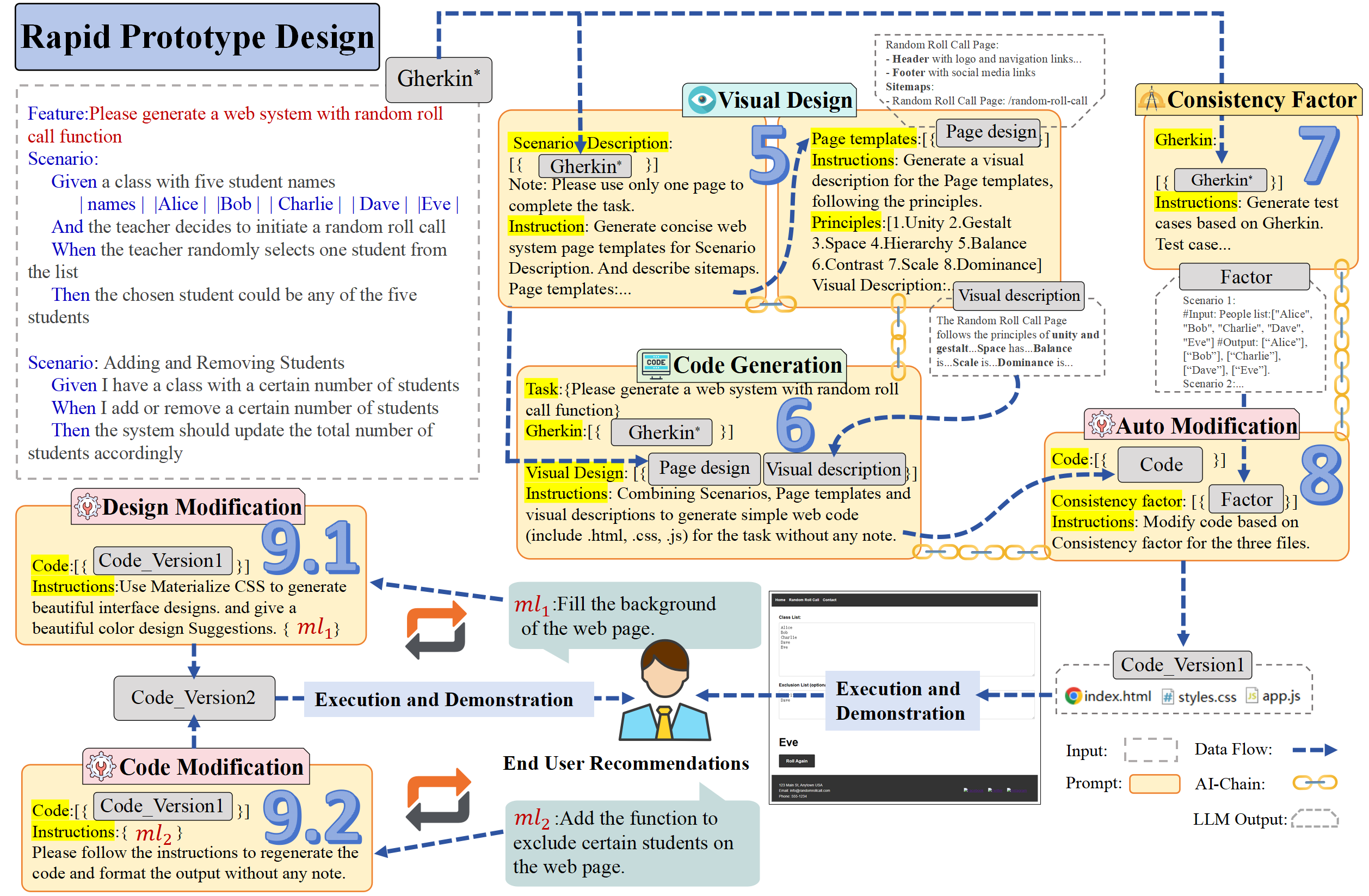}}
\caption{Overview Diagram of the Rapid Prototype Design Component. The numbers in the figure are the AgileGen execution sequences. The dotted arrows show the data flow of each component.}
\label{fig:scRapid}
\end{figure}

The Rapid Prototype Design component is tasked with generating software prototypes from Gherkin. The implementation details, as shown in Figure \ref{fig:scRapid}, follow the output $\mathrm{Gherkin^*}$ from the above Scenario Design component, proceeding through 5: Visual Design $\rightarrow$ 6: Code Generation $\rightarrow$ 7: Consistency Factor $\rightarrow$ 8: Auto Modification $\rightarrow$ End User  Acceptance and Recommendation Decision-making $\rightarrow$ 9.1: Design Modification / 9.2: Code Modification, to ultimately output the software's code. Here, 9.1 and 9.2 represent two interactive choices for modifications, namely visual design modifications and functional code modifications, respectively.
After the code is executed and presented to the end-user, they can choose and input the appropriate modification method based on their needs. The modified code will be executed and presented again for the end-user to make decisions. Notably, our method interacts with the end-user iteratively until the user accepts the generated result.

To ensure the integrity of the $\mathrm{Gherkin^*}$ file, we merge the separate scenarios generated by NL to Gherkin and prepend the file with the keyword ``Feature:$nl$" at the beginning.

\textbf{5: Visual Design}. Visual design is crucial for user-focused software. It combines text with graphics to capture attention and improve user experience. Even if end-users lack the ability of well-described UI/UX design, they still expect a well-designed interface. Visual design encompasses two parts: \textit{page design} and \textit{visual description}.

\textit{Page design} is a key part of the site organization in the research-based Web Design and Usability Guidelines~\cite{leavitt2006based[59]}. 
As shown on the left of Figure \ref{fig:scRapid}(5), this part designs the page content description for the website application through $\mathrm{Gherkin^*}$. Specifically, $\mathrm{Gherkin^*}$ is input into the ``Scenario Description:” and an instruction guides the large language models to generate a page design description. The generated result is shown in the $Page~design$ dashed expansion box in Figure~\ref{fig:scRapid}, including page content information such as Header, Form, Footer, etc. Particularly, Sitemaps describe the URL structure of the webpage.
The \textit{visual description}, as shown on the right of Figure~\ref{fig:scRapid}(5), takes $Page~design$ as input, with instructions and eight principles guiding the large language models to design the visual description. The incorporation of 8 successful visual design principles\footnote{https://www.usability.gov/what-and-why/visual-design.html} aims to guide the generation of descriptions centered around these principles. The generated result of the visual description is displayed in Figure~\ref{fig:scRapid} under \textit{Visual description}.
Our agent have generated a comprehensive visual description based on the page design. It outlines which content and functions should be emphasized.

\textbf{6: Code Generation}.
The purpose of code generation is to guide the large language models to generate code based on the Task, which stems from the end-user's raw requirement description. As shown in Figure~\ref{fig:scRapid}(6), the AI-Chain design for code generation includes Gherkin and Visual Design input interfaces. These interfaces respectively receive $\mathrm{Gherkin^*}$ containing human decisions and \textit{Page design \& Visual description} related to visual design. In practice, the large language models' generation results are diverse, and to facilitate the segmentation of code files, we also prescribe the starting format for generating code. Underneath the instructions, we add "Please generate the codes for the three files in <Code> without any note" and the following format (taking a website application as an example): 1.index.html:
```html
<Code>
```
end index.html
2.style.css:
```css
<Code>
```
end style.css 
3.script.js:
```javascript
<Code>
```
end script.js 
In this way, we can utilize the code file extraction rules described in the \textit{Acceptance and Recommendation Decision-making} section referred to in \ref{sec:interface}, to segment different code files. The generated code files are denoted by ``Code”.

\textbf{7: Consistency Factor}
In order to ensure a generated software meets end-user requirements requires a consistent factor to drive code generation consistent with user requirements. Although the one-time generated code may have good visual effects, it often lacks core functionality. This may be because the code generation part has input too many descriptions, causing the large language models to overlook the implementation of the most basic functions. To address this issue, we generate testable cases that are consistent with business logic from user stories in Gherkin. As shown in Figure \ref{fig:scRapid}(7), the AI-Chain design of the consistency factor consists of a Gherkin input interface and an instruction. The generated result is denoted by Factor, generating at least one business logic case for each scenario.
The cases provides sample inputs and outputs for the code modification part, facilitating the modification of functionality incorrectness.

\textbf{8: Auto Modification}. The Auto Modification part primarily modifies the code to better align with end-user requirements. The AI-Chain design is shown in Figure \ref{fig:scRapid}(8), where \textit{Auto Modification} automatically makes modifications to the "Code" after code generation based on the consistency factor. The AI-Chain design combines the consistency factor and the original code ("Code") to regenerate the code, denoted as Code\_Version1. This process aims to focus more on business logic cases, improving the functional correctness of the software code.

\textbf{9.1 \& 9.2: Design Modification and Code Modification}. Design Modification and Code Modification are related to the end-user's acceptance and recommendation decision-making choices. The execution result of ``Code\_Version1" is subject to the end-user's decision. When the end-user wishes to modify the interface design ($ml_1$), \textit{Design Modification} is selected. ``Code\_Version1" along with the user's decision opinion $ml_1$ are input into their respective input interfaces. When the end-user wishes to modify functionality ($ml_2$), \textit{Code Modification} is chosen. The end-user decision opinion $ml_2$ is input as an instruction, followed by modifications to ``Code\_Version1". The modified code ``Code\_Version2" can be executed again, allowing users to propose modifications, thereby implementing iterative acceptance with the end-user in this manner.


\textbf{Summary:} In the Scenario Design component, steps 1 to 4 involve the end-user in proposing requirements and making scenario decisions. From the end-user's raw requirement $nl$, a requirement with acceptance criteria, $\mathrm{Gherkin^*}$, is generated. This $\mathrm{Gherkin^*}$ serves as the input for the Rapid Prototype Design component, where steps 5 to 8 involve automatic prototype design by AI without end-user involvement. Steps 9.1 and 9.2 are triggered based on modification recommendations from the end-user, allowing them to choose different inputs and actions on the interface.
This approach places end-user involvement at both the beginning and end of the process, with AI handling the middle stages. By doing so, it maximizes the strengths of both end-users and AI, working together like pieces of a jigsaw puzzle to achieve generative software development.

\section{EXPERIMENTAL SETUP}
This section lists some research questions (RQs), followed by a description of the datasets used for the experiment and the evaluation metrics used to answer them.

\subsection{Research Questions}
In this paper, we introduce an Agile-based software development through the Human-AI Teamwork framework and collaborate with end-users to create personalized software.
There are some issues that interest us:
(1) We assess the effectiveness of software generated by our framework, including research on the quality of generated software and user satisfaction with the generation app.
(2) We analyze whether the high quality of generated software is due to the framework design or the base model performance, including ablation studies and case analyses.
(3) We evaluate user satisfaction with the interaction experience, including the number of interactions, waiting time, cost comparison, and surveys on user interaction experience.
Summary, we aim to address the following research questions:

\textbf{RQ1}: How effective are the software applications generated by our framework?

\textbf{RQ2}: What is the reason for the better performance of our framework?

\textbf{RQ3}: What is the human-computer interaction experience for end-users during the AgileGen process?

\subsection{Datasets}

\begin{table}[th]
\small
\caption{Displayed data for 30 website application cases in ``50projects50days".}
\begin{tabular}{llc}
\hline
Project Name & Link (https://50projects50days) & Description Words \\ \hline
07 Split Landing Page & .com/projects/split-landing-page/ & 41 \\
08 Form Wave & .com/projects/form-wave/ & 20 \\
09 Sound Board & .com/projects/sound-board/ & 18 \\
10 Dad Jokes & .com/projects/dad-jokes/ & 32 \\
11 Event KeyCodes & .com/projects/event-keycodes/ & 22 \\
12 Faq Collapse & .com/projects/faq-collapse/ & 39 \\
16 Drink Water & .com/projects/drink-water/ & 36 \\
17 Movie App & .com/projects/movie-app/ & 46 \\
18 Background Slider & .com/projects/background-slider/ & 29 \\
19 Theme Clock & .com/projects/theme-clock/ & 29 \\
22 Drawing App & .com/projects/drawing-app/ & 32 \\
23 Kinetic Loader & .com/projects/kinetic-loader/ & 23 \\
24 Content Placeholder & .com/projects/content-placeholder/ & 42 \\
25 Sticky Navbar & .com/projects/sticky-navbar/ & 29 \\
27 Toast Notification & .com/projects/toast-notification/ & 38 \\
28 Github Profiles & .com/projects/github-profiles/ & 29 \\
30 Auto Text Effect & .com/projects/auto-text-effect/ & 38 \\
31 Password Generator & .com/projects/password-generator/ & 29 \\
32 Good Cheap Fast & .com/projects/good-cheap-fast/ & 60 \\
33 Notes App & .com/projects/notes-app/ & 5 \\
39 Password Background & .com/projects/password-strength-background/ & 23 \\
41 Verify Account Ui & .com/projects/verify-account-ui/ & 37 \\
42 Live User Filter & .com/projects/live-user-filter/ & 37 \\
43 Feedback Ui Design & .com/projects/feedback-ui-design/ & 41 \\
44 Custom Range Slider & .com/projects/custom-range-slider/ & 36 \\
46 Quize App & .com/projects/quiz-app/ & 29 \\
47 Testimonial Box Switcher & .com/projects/testimonial-box-switcher/ & 33 \\
48 Random Image Feed & .com/projects/random-image-feed/ & 22 \\
49 Todo List & .com/projects/todo-list/ & 30 \\
50 Insect Catch Game & .com/projects/insect-catch-game/ & 55 \\ \hline
\end{tabular}
\label{tab:50day}
\end{table}

\begin{table}[]
\small
\caption{Ten reality projects created by end-users.}
\begin{tabular}{ll}
\hline
\multicolumn{1}{c}{Project Name} & \multicolumn{1}{c}{Descriptions} \\ \hline
Bookkeeping Assistant & I need a bookkeeping assistant website. \\
Random Roll Call & Please generate a web system code with random roll call function. \\
Shopping Site & I need a shopping website. \\
Gomoku & Please design a basic Gomoku game. \\
Draw Flowers & Please design a website that can draw different types of flowers. \\
Weather Forecast & \begin{tabular}[c]{@{}l@{}}A weather display interface, which can show the weather condition of \\ city and future weather report.\end{tabular} \\
Online Timer & \begin{tabular}[c]{@{}l@{}}Provides a simple interface where users can set the duration of the \\ timer and start or stop the timer.\end{tabular} \\
Currency Converter & Please generate a currency converter webpage. \\
Online Translator & Please generate an online translator website. \\
Event Reminder & I would love to have an website that have add event reminder function. \\ \hline
\end{tabular}
\label{tab:10data}
\end{table}

The datasets "50projects50days" and "SRDD" are used to evaluate our method.
We first use a modified ``50projects50days" dataset to evaluate website application development capability and code generation performance.
To evaluate the effectiveness of our framework, we test it using the ``50projects50days\footnote{https://github.com/bradtraversy/50projects50days}" open-source project on GitHub. Each web project comprises three files (.html, .css, and .js) that determine the project's layout, style, and interaction logic. These code files serve as reference codes for evaluating the quality of the generated code.
We excluded 20 projects of simple visual animation types and evaluated a total of 30 projects. As shown in Table \ref{tab:50day}, we present the 30 projects used for evaluation. The number before each project name represents its ID in the ``50projects50days" dataset. ``Description Words" indicates the number of natural language task words describing the project, often called the task description. We automatically construct task descriptions to reduce human bias by inputting each project's frontend interface into a multimodal ChatGPT to generate a content description.
Moreover, we asked participants to create their web application requirements to prevent the likelihood that the large language models had previously been trained on GitHub, including ten user-created projects. As shown in Table~\ref{tab:10data}, we present the names and descriptions of ten reality projects created by users. These descriptions exhibit characteristics that are more concise, conversational, and vague.

\begin{table}[]
\caption{From a random sample of 20 software projects in the SRDD dataset.}
\small
\label{tab:SRDD}
\begin{tabular}{cll|cll}
\hline
ID & \multicolumn{1}{c}{Project Name} & \multicolumn{1}{c|}{Category} & ID & \multicolumn{1}{c}{Project Name} & \multicolumn{1}{c}{Category} \\ \hline
1 & SportArena & Game-Sport Game & 11 & SecureGuard & Work-Security \\
2 & Shot\_Accuracy\_Trainer & Game-Sport Game & 12 & Virus\_Protector & Work-Security \\
3 & Strategic\_Alliance & Game-Strategy Game & 13 & RecommendationMate & Life-Personalization \\
4 & Battle\_Masters & Game-Strategy Game & 14 & Mindful\_Meditation & Life-Personalization \\
5 & News\_Viewer & Education-News & 15 & ConnectionHub & Life-SocialNetwork \\
6 & NewsMeter & Education-News & 16 & SocialMatch & Life-SocialNetwork \\
7 & Storytime\_Fun & Education-Family\&Kids & 17 & Scene\_Detection & Creation-Video \\
8 & Time\_Travel\_Adventure & Education-Family\&Kids & 18 & VideoClipper & Creation-Video \\
9 & Business\_Analytics & Work-Business & 19 & Vector\_Creator & Creation-Graphics \\
10 & Support\_Ticket\_System & Work-Business & 20 & ZoomSketch & Creation-Graphics \\ \hline
\end{tabular}
\end{table}

Second, we use the ChatDev dataset SRDD~\cite{qian2024iterative,qian2024scaling}. This dataset consists of 1200 software task prompts in the form of Name/Description/Category and does not include reference code. The software task prompts in SRDD are divided into five main categories: Education, Work, Life, Creation, and Game. These main categories are further subdivided into 40 different subcategories, each containing 30 unique tasks. Each task describes more complex and diverse software requirements. We perform stratified sampling of the dataset by software category, randomly selecting two subcategories from the 40 different subcategories within each main category and sampling two unique task prompts from each subcategory. The sampled dataset contains 20 projects, as shown in Table~\ref{tab:SRDD}.

\subsection{Implementation details}
Our method, AgileGen(3.5), uses gpt-3.5-turbo as its base model, while AgileGen(4) uses gpt-4-1106-preview. These are autoregressive language models based on generative pre-trained transformers developed by OpenAI. The latest large language model released by OpenAI, gpt-4o-2024-05-13, will also be considered for comparative experiments in RQ1.1, representing the state-of-the-art results in large language models.
To stabilize the framework automation process, default values were used for the control GPT output confidence parameter "Top-P" set to 1.0 and the Frequency and Presence penalties set to 0.0. The control output randomness parameter "temperature" was set to 0.3, and the "max token size" was set to 4096 due to the long generated code. 
The maximum number of generated scenarios is set to 10. In the memory pool, the similarity threshold is set to 0.7.
Additionally, considering that some baseline methods are not designed to generate web code, which might result in a lower code executability rate, we will append "Web code (.html, .css, .js) is preferred" to the task descriptions of the baselines during the experiments. This adjustment aims to increase the likelihood of the baseline methods generating web code, thereby making the comparative experiments more fair.

\section{Experiment Results}

\subsection{\textbf{RQ1:How effective are the software applications generated by our framework?}}
\label{subsection:RQ1}
Generative software development agents aim to create software that align with the end-user's requirements. We compare AgileGen with existing methods regarding the quality of code generation, functional completeness, and user experience. The goal is to investigate whether our approach offers superior output results to other generative software development agents, particularly in generating complete software projects. 
\subsubsection{RQ1.1: How is the code quality of the software applications generated by our framework according to the automatic evaluation metrics?}
\label{RQ1.1}
\paragraph{\textbf{Motivation}}
Generative software development agents aim to develop code that is highly quality and functionally complete. Automatic evaluation can assess the code generated by different methods from objective perspectives, particularly in test case pass rates, code similarity, time, and cost. This testing will demonstrate the comparative effectiveness of different methods across various projects.
\paragraph{\textbf{Methodology}}
To explore RQ1.1, we will introduce the approach step by step according to a. Automatic evaluation metrics, b. Baselines, and c. Experimental Setup.

a. Automatic Evaluation Metrics. We use the automatic evaluation metrics CodeBLEU and Pass@k to assess code quality. Additionally, we use Time (s), Pro\_Code, and Cost (\$) metrics to evaluate the time cost, code token proportion, and monetary cost when different generative software development agents generate software applications. CodeBLEU~\cite{DBLP:journals/corr/abs-2009-10297[10]2}: An automatic evaluation metric designed for code generation, assessing aspects including syntax, semantics, data flow, and code structure. CodeBLEU incorporates the advantages of BLEU in n-gram matching and further infuses code syntax and semantics through the Abstract Syntax Tree (AST) and data flow.
Pass@k~\cite{DBLP:conf/emnlp/0034WJH21[19]} unbiased version, evaluates the functional correctness of the generated top-k code by running test cases. This paper primarily focuses on the evaluation of Pass@1, as we consider the ability of a generative software development agent to correctly generate code in a single attempt to be important. Pass@k is described as follows:
\begin{equation}
\text { Pass } @ \mathrm{k}=\mathbb{E}_{\text {Problems }}\left[1-\frac{\left(\begin{array}{c}
n-c \\
k
\end{array}\right)}{\left(\begin{array}{c}
n \\
k
\end{array}\right)}\right]
\end{equation}
Time (s): Represents the waiting time during the system execution. The metric only includes the time automatically executed by the system. Cost (\$) focuses on the tokens consumed by the agent. We use the uniform token calculation method provided by large language models' tokenizer\footnote{https://github.com/huggingface/tokenizers} and calculate the average Cost.
Proportion of code tokens (Pro\_Code) is the ratio of code tokens generated by the generative software development agent to the total tokens, as counted by the tokenizer method, described as follows:
\begin{equation}
\begin{gathered}
 \text{Pro\_Code} =\frac{Code~ tokens }{ Total~ tokens } \\
Total~ tokens= Input~ tokens+ Output~ tokens
\end{gathered}
\end{equation}

b. Baselines.
We use the "50projects50days" dataset to evaluate our method against software applications generated by seven state-of-the-art methods, including AutoGPT, ChatDev\_23 (an early version of ChatDev from 2023), ChatDev\_24 (the latest version of ChatDev), GPT-Engineer, GPT3.5+CoT, GPT4+CoT, and GPT4o+CoT.

\begin{itemize}
    \item The SOTA auto-thinking Agent, AutoGPT~\cite{AutoGPT[32]}, is based on ChatGPT, which is a pre-trained transformer (GPT). AutoGPT acts autonomously without requiring human agents to prompt every action. It is available on GitHub but requires programming experience, as it runs on Python and necessitates OpenAI and Pinecone API keys.
    \item The SOTA questioning Agent, GPT-Engineer~\cite{Gpt-engineer[4]2}, is a GitHub open-source project utilizing GPT-4 to automate the software engineering process. It operates by accepting a requirement and then posing questions back to the user. Similarly, this project requires some programming skill for environment setup and operation.
    \item The SOTA multi-agents collaborative Agent, ChatDev\_23~\cite{DBLP:journals/corr/abs-2307-07924[1]2}, builds on ChatGPT by assigning different roles to agents, with agents of various roles automatically fulfilling user requirements through a chat chain as the facilitator. ChatDev\_24~\cite{qian2023experiential,qian2024iterative,qian2024scaling} adds the acquisition and use of historical experience based on previous foundations.
    \item Chain-of-Thought (CoT) retains the "Code Generation" part of our framework, replacing the rest with the task plus ``Let's think step by step." It guides the large language model to decompose the user's requirements and generate code gradually. The experiments are based on GPT3.5-Turbo (GPT3.5 + CoT), GPT4-Turbo (GPT4 + CoT) and GPT4o (GPT4o + CoT) respectively.
\end{itemize}

c. Experimental Setup.
In the ``50projects50days" dataset, we generate the test cases for the JavaScript code using Github Copilot, based on the reference code (with the prompt: "Generate test code with test cases for this code."). We will adjust these test cases to ensure they are reasonable and include boundary testing. The pass rate of these test cases evaluates the functionality completeness of the generated software code.
Additionally, to obtain the generated software applications code fairly, we will use execution scripts to automatically run nine generative software development methods based on the task descriptions in the "50projects50days" dataset. We will record the generated software codes and their spend time and cost. Generating end-user applications in this experiment will exclude human expertise and intelligence to ensure fairness.
\paragraph{\textbf{Result Analysis}}

\begin{figure}[t]
\centerline{\includegraphics[width=\textwidth]{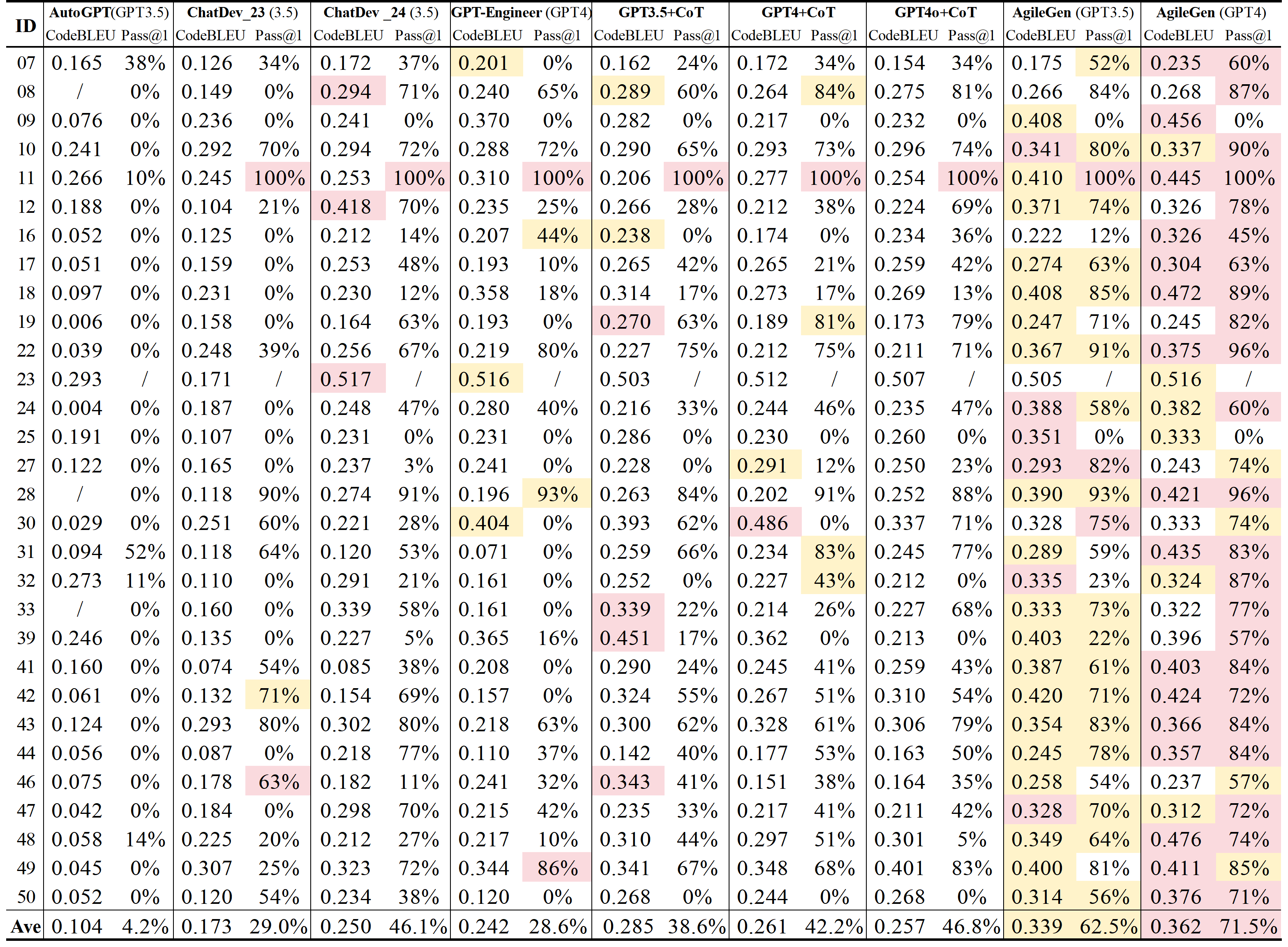}}
\caption{Automatic metrics CodeBLEU and Pass@1 are used to evaluate our method and existing methods on the "50projects50days" dataset (Thirty functionally complex projects were retained for evaluation.) Among these, Our AgileGen(GPT4), GPT-Engineer and GPT4+CoT use GPT4-Turbo as the base model, while AutoGPT, ChatDev (ChatDev\_23, ChatDev\_24), GPT3.5+CoT, and our AgileGen(GPT3.5) use GPT-3.5 as the base model. Additionally, GPT4o+CoT is based on the latest GPT4o model. Red indicates the best results, and yellow signifies the second-best results. The ID corresponds to the project number in the dataset.}
\label{fig:autotable}
\end{figure}

We use automatic metrics to evaluate the quality of software generation by our method and existing methods. Figure~\ref{fig:autotable} shows the results on the ``50projects50days" dataset. The red area indicates the best results on the project, and the yellow area indicates the second-best results. The `/' symbol denotes an inability to compute, where its appearance under the CodeBLEU metric implies that the method tried multiple times but could not generate complete code for the project. For the Pass@1 metric, the appearance of '/' for project ID23 indicates it does not contain JavaScript code. Hence, no method could compute the metric value for it.

As shown in Figure~\ref{fig:autotable}, our AgileGen achieved the best average results on both metrics. AgileGen (GPT3.5) results are 0.339 and 62.5\% in CodeBLEU and Pass@1, better than the state-of-the-art ChatDev\_24 (GPT3.5) by 0.089 and 16.4\% in CodeBLEU and Pass@1. AgileGen (GPT4) results are 0.362 and 71.5\%, outperforming the latest GPT4o model by 0.105 and 24.7\% in CodeBLEU and Pass@1. Overall, our AgileGen method, based on GPT-4 and GPT-3.5 models, dominated 73.3\% of the best results and 83.3\% of the second-best results in the CodeBLEU metric. In the Pass@1 metric, our method achieved the best results in 83.3\% and the second-best in 70\%.
For both ID09 and ID25, the Pass@1 metric is 0.0\% across all methods. This is because ID09 is a "Sound Board" task, where many of the multimedia resource URLs generated by agents were unverified, leading to many not loading or being missing, thus failing the test cases. ID25 requires completing a ``Sticky Navbar" task, where the dataset's JavaScript code implements a color change in the navbar as one scroll down. Still, all methods fail to generate the corresponding JS code while generating the page code. This might relate to the task description. In these cases, our method still achieved the best results in CodeBLEU. ID23 is a simple ``Kinetic Loader" loading symbol display, where our method, along with GPT-Engineer, ChatDev, and all CoT methods, could all generate the functionality, but the rotation style of the loading symbol differed from the reference code, showcasing the diversity of the generation frameworks. 
ID16 is a project that records daily water intake (``Drink Water''). Due to the diversity of implementation methods, our AgileGen (3.5) adopts an increasing or decreasing method to meet the daily water limit. Projects implemented with the decreasing method are less likely to pass test cases, yet our AgileGen (4) achieved the best results in the ID16 project.
ID31 requires the functionality of a password generator (``Password Generator''). Our AgileGen(3.5) sometimes causes the generated passwords to appear on a redirect page, which users cannot return due to the page design. However, this issue is mitigated when our method is based on the GPT-4 model. The inclusion of Sitemaps in page design guides GPT-4 to generate code that redirects to specified pages. In this case, Sitemaps generally do not generate multiple links when only a single function is involved.
On project ID30, ``Auto Text Effect," our method was slightly inferior in the CodeBLEU metric compared to other methods but achieved the best result in Pass@1. This is because the task required implementing a webpage with an interactive feature that displays text characters one by one, while other methods tended to be implemented at the word level. 

\begin{table}[]
\caption{Comparison of execution performance for different methods on the ``50projects50days", including average time (Time(s)), the proportion of code tokens (Pro\_Code), and cost (Cost(\$)).}
\label{tab:system}
\begin{tabular}{lccc}
\hline
 & Time(s) & Pro\_Code & Cost(\$) \\ \hline
AgileGen (GPT3.5) & 68.625 & 11.57\% & 0.0110\$ \\
AgileGen (GPT4) & 117.700 & 14.76\% & 0.1490\$ \\
AutoGPT (GPT3.5) & 213.325 & 0.28\% & 0.0442\$ \\
ChatDev\_23 (GPT3.5) & 343.875 & 1.91\% & 0.0259\$ \\
ChatDev\_24 (GPT3.5) & 273.350 & 2.16\% & 0.0192\$ \\
GPT-Engineer (GPT4) & 118.875 & 8.25\% & 0.1436\$ \\
GPT3.5+CoT & 26.550 & 19.71\% & 0.0028\$ \\
GPT4+CoT & 89.825 & 22.26\% & 0.0819\$ \\
GPT4o+CoT & 32.350 & 20.94\% & 0.0590\$ \\ \hline
\end{tabular}
\end{table}

Then, we explore the system performance of different methods, as shown in Table~\ref{tab:system}. 
\begin{itemize}
    \item \textbf{Time}, measured in seconds, reflects the average duration for system execution. Our system generates a software in about one minute, while other methods have longer waiting times. CoT-related methods show shorter waiting times due to fewer input-output operations. However, our method's generation effect is significantly improved compared to methods like CoT, and the duration of AgileGen (GPT3.5) is still shorter than GPT-4+CoT.
    \item  \textbf{Pro\_Code} indicates the proportion of generated code tokens to total tokens; a higher ratio reflects fewer intermediate products. AgileGen has the smallest proportion of intermediate products, except for the CoT method, showing that our method does less generate redundant intermediates while achieving the best results.
    \item \textbf{Cost} represents the average expenditure per project, measured in US dollars. Our method is more cost-efficient, saving nearly two times per project generation compared to ChatDev and four times per project compared to AutoGPT.
\end{itemize}

\subsubsection{RQ1.2: How effective is the software applications generated by our framework in meeting user requirements?}
\label{RQ1.2}
\paragraph{\textbf{Motivation}}
Automatic metrics evaluate software applications generated by different methods from a functional perspective. However, non-functional requirements, such as visual appeal, need to be better assessed. The ultimate goal of generative software development methods is to develop software applications that align with user requirements. Therefore, human evaluation, including non-functional aspects, better verifies whether the generated software applications meet user requirements. Additionally, to explore the robustness of our method, the SRDD dataset~\cite{qian2024iterative} is introduced for evaluation. Human evaluation is necessary for this dataset since it does not contain reference code and cannot be evaluated using automatic metrics.

\paragraph{\textbf{Methodology}}
To explore RQ1.2, we will introduce the approach step by step according to: a. Human Evaluation Metrics, b. Baselines, and c. Experimental Setup.

a. Human Evaluation Metrics. We use human metrics, Code Executability~\cite{DBLP:journals/corr/abs-2308-00352[2]2} and User Experience Questionnaire (UEQ)~\cite{schrepp2019all[60]}, to evaluate the non-functional requirements of the generated software applications.

Code Executability~\cite{DBLP:journals/corr/abs-2308-00352[2]2} evaluates from a human perspective whether the software application is a complete failure or meets the task specifications to some extent.
The assessment is based on levels ranging from a complete failure of ``0" to perfect execution and full compliance with task specifications of ``3". The definition of levels is as follows:
\begin{itemize}
    \item 0 point: The code is non-functional or deviates completely from the requirements.
    \item 1 point: The code executes, but it may not meet all workflow requirements.
    \item 2 point: The code executes and aligns mostly with the requirements.
    \item 3 point: The code functions flawlessly, and the output corresponds exactly to the specifications.
\end{itemize}

User Experience Questionnaire (UEQ): User experience assesses the end-user's subjective evaluation of generated website applications. Since our static website applications don't involve back-end or server aspects, traditional website evaluation scales such as "responsiveness" and "navigation experience" are not applicable (e.g., Wammi, SUPR-Q). The UEQ~\cite{schrepp2019all[60]} (User Experience Questionnaire) offers an efficient method to evaluate interactive products, with 26 questions scored from -3 to 3. Six user experience metrics are derived from these scores: Attractiveness, Perspicuity, Efficiency, Dependability, Stimulation, and Novelty, reflecting the overall performance of static website applications.

b. Baselines. The baselines on the "50projects50days" dataset include seven different methods: AutoGPT~\cite{AutoGPT[32]}, ChatDev\_23~\cite{DBLP:journals/corr/abs-2307-07924[1]2}, ChatDev\_24~\cite{qian2023experiential,qian2024iterative,qian2024scaling}, gpt-engineer~\cite{Gpt-engineer[4]2}, GPT3.5+CoT, GPT4+CoT, and GPT4o+CoT. For details, refer to the RQ1.1 section~\ref{RQ1.1}.
The baselines on the "SRDD" dataset include ChatDev\_24 (GPT4) and ChatDev\_24 (GPT3.5).
\begin{itemize}
    \item ChatDev\_24 (GPT3.5) is consistent with the ChatDev\_24 used on the ``50projects50days" dataset.
    \item ChatDev\_24 (GPT4) is a version of ChatDev\_24 based on GPT-4 Turbo, obtained by modifying the configuration parameter `model'.
\end{itemize}

c. Experimental Setup.
\textbf{Participants}: To conduct a human evaluation of generative software development methods, we designed a survey to collect feedback from participants with various professional backgrounds. First, we screened individuals from different industries on professional social platforms. Then, we designed a detailed email invitation explaining the survey's purpose and the tasks participants needed to complete. We also offered a small financial compensation of \$20 per person to increase response rates. We invited 20 participants to help us evaluate the software applications generated by different methods. Among these participants were 13 males, six females, and 1 of unknown gender. 45\% came from various professions, including Consulting Media, Design Industry, Education, Finance, Medical Treatment, Service, and Online Retailers. 55\% were students from other institutions, with five majoring in software engineering, 3 in computer science and technology, and 3 in new media and communication. Regarding age distribution, 60\% were aged 18\~{}24, 35\% were aged 25\~{}34, and 5\% were aged 35\~{}44. Among the 20 participants, 8 had only basic computer operation knowledge, 6 were novice programmers (0\~{}1 year of experience), 4 were beginners (1\~{}3 years of experience), and 2 were experienced programmers (more than three years of experience). These participants had never been involved in any stage of our method's development, nor had they used our method to develop any software applications, ensuring their evaluations were objective and unbiased.

\textbf{Study procedure}: To ensure that all participants understand the rules for scoring Code Executability and UEQ and how to execute the generated web applications, we organized an online training session. During this session, we demonstrated the execution and scoring process using two web applications that were not part of the evaluation dataset. Following the demonstration, we conducted a Q\&A session to address any questions. Additionally, we asked 10 participants from different professions to propose their own web application requirements, which we collected as part of the "50projects50days" dataset. After the training session, we used scripts to automatically generate web applications based on the newly proposed requirements. There are a total of 60 projects to be evaluated across the "50projects50days" and "SRDD" datasets. Participants who proposed web application requirements will be assigned their own project and two random projects, while other participants will be randomly assigned three projects. Assigned projects will not be reassigned to ensure no duplication. The method name in each project will be anonymized, retaining only a method ID, and will include a Code Executability form and a UEQ scoring form corresponding to that ID. For fairness, the web applications used for evaluation of AgileGen are generated through scripts with no human interaction during the generation process.

\paragraph{\textbf{Result Analysis}}

\begin{figure}[h]
\centerline{\includegraphics[width=0.7\textwidth]{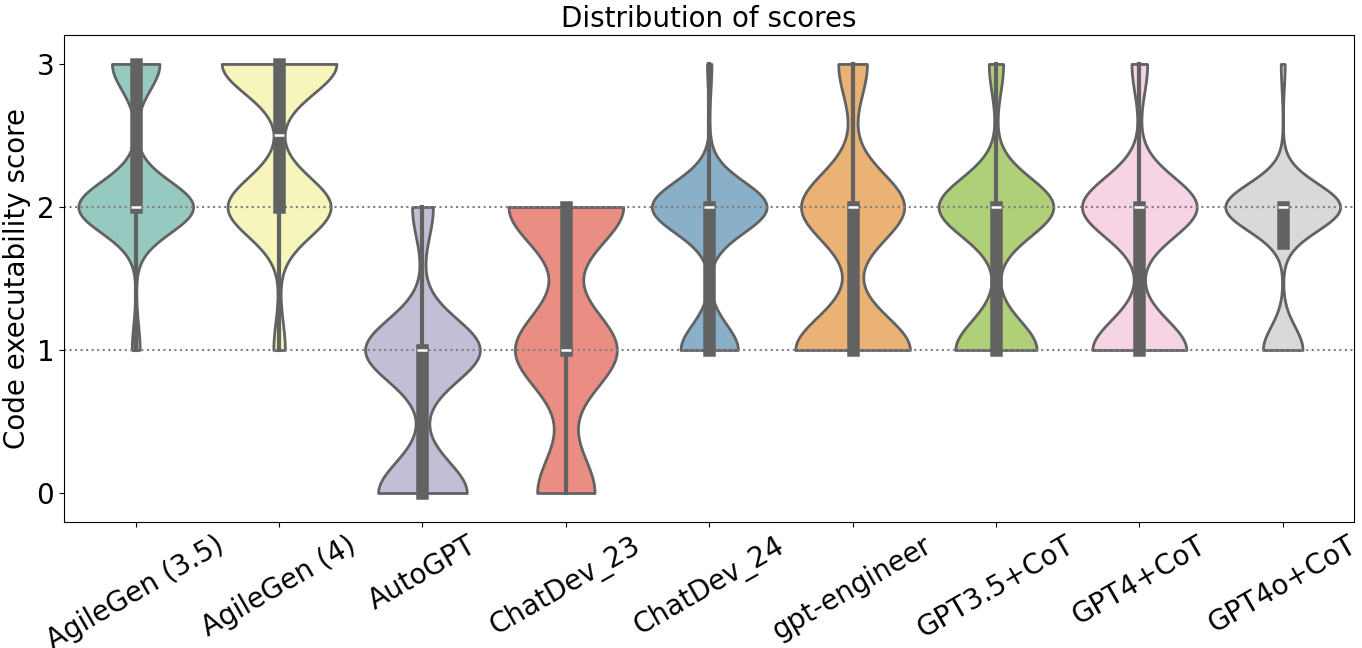}}
\caption{A comparison of the distribution of human evaluation metric Code executability scores across 40 projects (including 10 real software cases) for different methods. The wider the distribution at a certain score, the higher the frequency of that score occurring, and vice versa.}
\label{fig:violin}
\end{figure}

First, We explore performance on the human evaluation metric of Code executability on ``50projects50days''. This metric reflects the functional quality of the generated code, specifically its consistency with the task description. As shown in the Figure~\ref{fig:violin}, scores for our method, AgileGen scored 2 in 45\% of the tasks and 3 in 50\%. For AutoGPT, the scores of 0 and 1 together account for 97.5\%, which may be due to the majority of the generated code being incomplete. ChatDev\_23 scores are primarily 1 and 2, accounting for 37.5\% and 42.5\%, respectively. The scores of ChatDev\_24 are mainly concentrated at 2, accounting for 65\%, with a small proportion of 3 scores at 2.5\%. This is significantly better than ChatDev\_23 across the entire score distribution. The performance of the GPT3.5+CoT method is higher than AutoGPT, which could be because the waterfall model they adopt may easily propagate errors from one agent to the next, making it difficult to generate software products that meet task descriptions. The distribution for ChatDev\_24, GPT-Engineer, GPT3.5+CoT, GPT4 + CoT and, GPT4o + CoT in the figure is similar. In comparison, AgileGen scored 2 in 45\% of the tasks and 3 in 50\%. Our method achieves a significantly higher proportion of 3 scores, outperforming ChatDev\_24 by 47.5\%, indicating that a process with a formal user story makes more detailed user requirements that are good for code generation. Our methods guided by an Agile methodologies with a Gherkin language, received higher scores, demonstrating that our method can improve the consistency of generated code with the task description.

\begin{figure}[h]
\centerline{\includegraphics[width=0.8\textwidth]{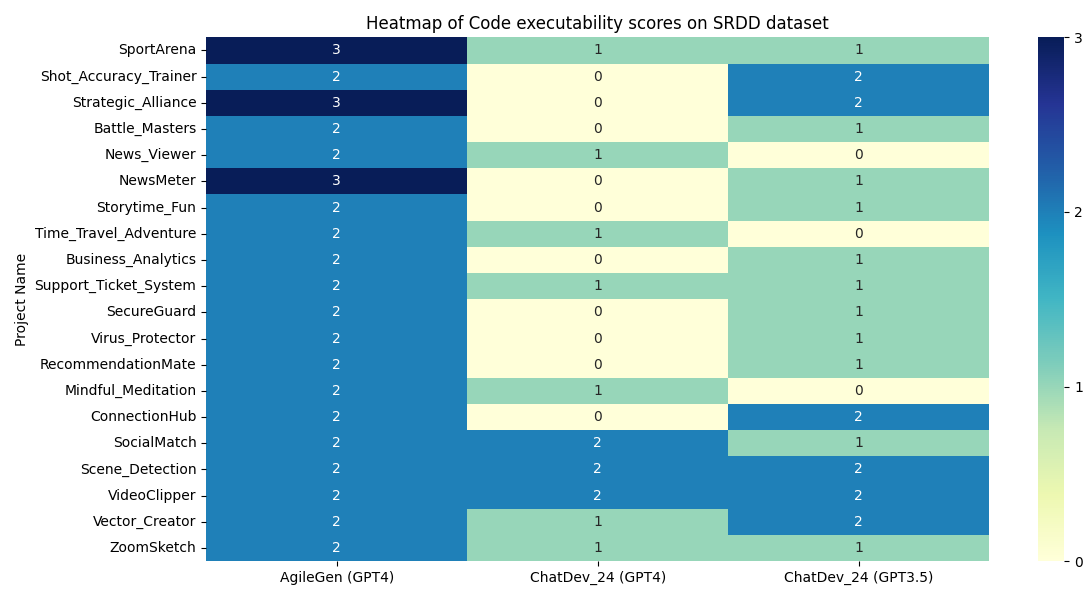}}
\caption{A comparison of the human evaluation metric Code executability scores distribution across 20 projects from the SRDD dataset. The closer the color is to blue, the higher the score, and the closer the color is to yellow, the lower the score.}
\label{fig:heatmap_SRDD}
\end{figure}

Second, to evaluate the robustness of our method, we utilized Code Executability to test its performance on the "SRDD" dataset, which contains more diverse and complex tasks. As shown in Figure~\ref{fig:heatmap_SRDD}, we compared the AgileGen and ChatDev methods on this dataset. Due to potential internal compatibility issues that resulted in poorer performance with the GPT-4-based ChatDev\_24, we used the more stable GPT-3.5-based ChatDev\_24 results as a reference. In the figure, darker blue indicates higher scores, while yellow indicates lower scores. We automatically generated 20 software projects for fairness using different methods without human interaction. ChatDev\_24 (GPT-4), 50\% of the tasks scored 0, 35\% scored 1, and 15\% scored 2, indicating that many software applications it generated faced compilation issues. Analysis revealed that most errors were due to syntax errors (e.g., undeclared variables) and missing custom package references in the main function. ChatDev\_24 (GPT-3.5), 15\% scored 0, 55\% scored 1, and 30\% scored 2, indicating that most generated applications were executable but only partially met the requirements. The main reason for scoring 1 was the incomplete or missing features in the generated GUI, such as the "Business\_Analytics" project displaying only a single text message without interactive operations and the "Storytime\_Fun" project having two unresponsive buttons.

In contrast, for AgileGen, 85\% of the tested projects scored 2, and 15\% scored 3, indicating that the applications it generated were executable and largely met the requirements, though rarely perfectly. For example, the "SportArena" project~\footnote{https://github.com/HarrisClover/AgileGen-SRDD-Experiment/tree/main/SportArena/AgileGen/html} scored 3, as shown in Figure~\ref{fig:case_SRDD}, while the "Shot\_Accuracy\_Trainer"~\footnote{https://github.com/HarrisClover/AgileGen-SRDD-Experiment/tree/main/Shot\_Accuracy\_Trainer/AgileGen/html} as shown in Figure~\ref{fig:interaction_case_3} scored 2. The task prompt for "SportArena" was "Develop a user-friendly software application that allows users to create and customize virtual sports arenas." Our method produced a web application for customizing features like the type and size of arenas and outputting configuration information to meet the task requirements. In contrast, the "Shot\_Accuracy\_Trainer" task prompt was more detailed, focusing on "using a virtual interface to practice shooting." Our method created a web application with functionalities such as "choosing a sport," "inputting distance and target size," "generating a shooting accuracy report," and "providing technical feedback." However, the "Start Training Session" button did not reflect the game's fun aspect. During feedback interviews, one participant said, "It feels more like a training recording web application than a game." Thus, achieving a perfect match with user requirements necessitates considering both functional and non-functional requirements. Individual non-functional requirements can vary, emphasizing the importance of human-computer interaction in our method.

\begin{figure}[h]
\centerline{\includegraphics[width=0.5\textwidth]{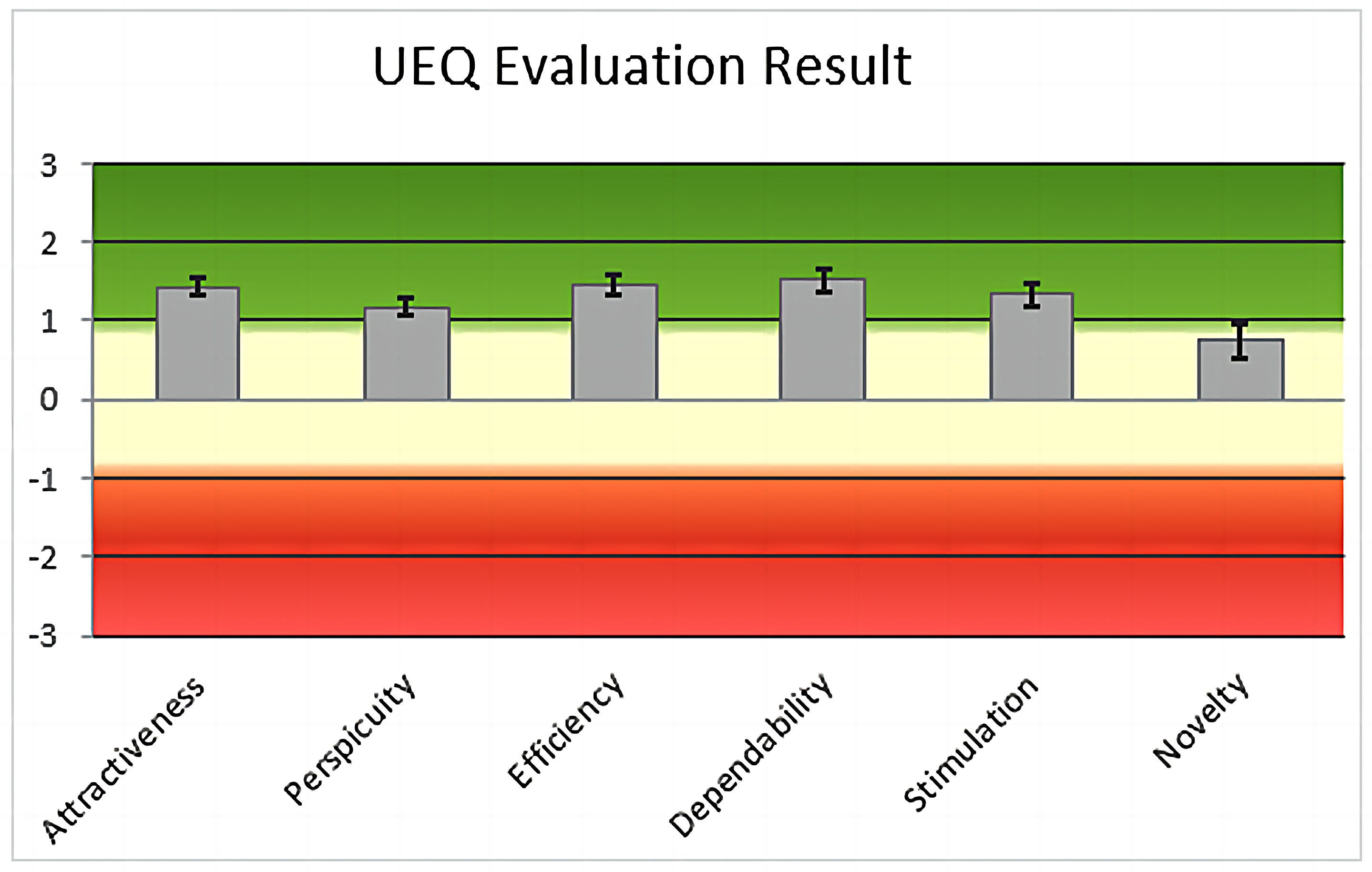}}
\caption{The average score of the UEQ evaluation on the six indicators. Red to green represents scores from low to high.}
\label{fig:UEQ}
\end{figure}

Finally, to test user satisfaction with the generated website applications, we utilized the User Experience Questionnaire (UEQ) on the ``50projects50days" dataset, and the results are depicted in Figure~\ref{fig:UEQ}. This metric primarily evaluates websites from a human subjective perspective, including non-functional requirements such as page aesthetics.
The metrics indicate that the software performs above average, with an overall average score of 1.26. All metrics have an average score above +1, except for innovation, which falls below +1. The "Attractiveness" and "Dependability" metrics performed exceptionally well, indicating that our framework meets end-user expectations and results in high satisfaction levels. However, the low innovation score may be attributed to our framework's reliance on large language models, which tend to generate common code on which they have been trained. Consequently, some web application designs lean towards being "conservative." Despite this, our scenario generation module still completes user requirements, offering unexpected features and resulting in positive scores for innovation. For instance, the participant who proposed the "Random Roll Call" project mentioned: "When I proposed the random roll call task, I just wanted a student list that could randomly select students to answer questions in class. I was surprised that the project also included an exclusion list, allowing me to exclude absent students."

\subsubsection{RQ1.3: How effective is the software applications generated by our framework in case studies?}
\paragraph{\textbf{Motivation}}
Case studies provide a detailed showcase of the generation effectiveness for individual projects, making it more intuitive than automatic or human evaluation metrics. Especially in terms of visual effects, case studies offer a more intuitive demonstration of the different outcomes produced by different methods.

\paragraph{\textbf{Methodology}}
For the exploration of RQ1.1, we will introduce the approach step by step according to: a. Baselines, b. Study Setup.

a. Baselines. In the case studies, we showcase six different methods on the "50projects50days" dataset: ChatDev\_23~\cite{DBLP:journals/corr/abs-2307-07924[1]2}, ChatDev\_24~\cite{qian2023experiential,qian2024iterative,qian2024scaling}, gpt-engineer~\cite{Gpt-engineer[4]2}, GPT3.5+CoT, GPT4+CoT, and GPT4o+CoT. On the "SRDD" dataset, our baselines include ChatDev\_24 (GPT4) and ChatDev\_24 (GPT3.5). For detailed information, please refer to section RQ1.2~\ref{RQ1.2}.

b. Study Setup. We selected case studies based on participants' Code Executability metric evaluations, focusing on projects with significant evaluation differences. In the "50projects50days" dataset, we chose the "bookkeeping assistant" project created by a participant for case analysis. Selecting a participant-created project helps avoid the risk of using code that large language models may have already been trained on from GitHub, showcasing the ability of different methods to create personalized projects for users. In the "SRDD" dataset, we selected three projects—"SportArena," "NewsMeter," and "VideoClipper"—for analysis, starting with those with the most significant differences in Code Executability scores. We have prepared two replication packages for public validation of the experiment on the "50projects50days" dataset~\footnote{https://github.com/HarrisClover/AgileGen-50projects50days-Experiment} and the "SRDD" dataset~\footnote{https://github.com/HarrisClover/AgileGen-SRDD-Experiment}.

\paragraph{\textbf{Result Analysis}}
\begin{figure}[h]
\centerline{\includegraphics[width=\textwidth]{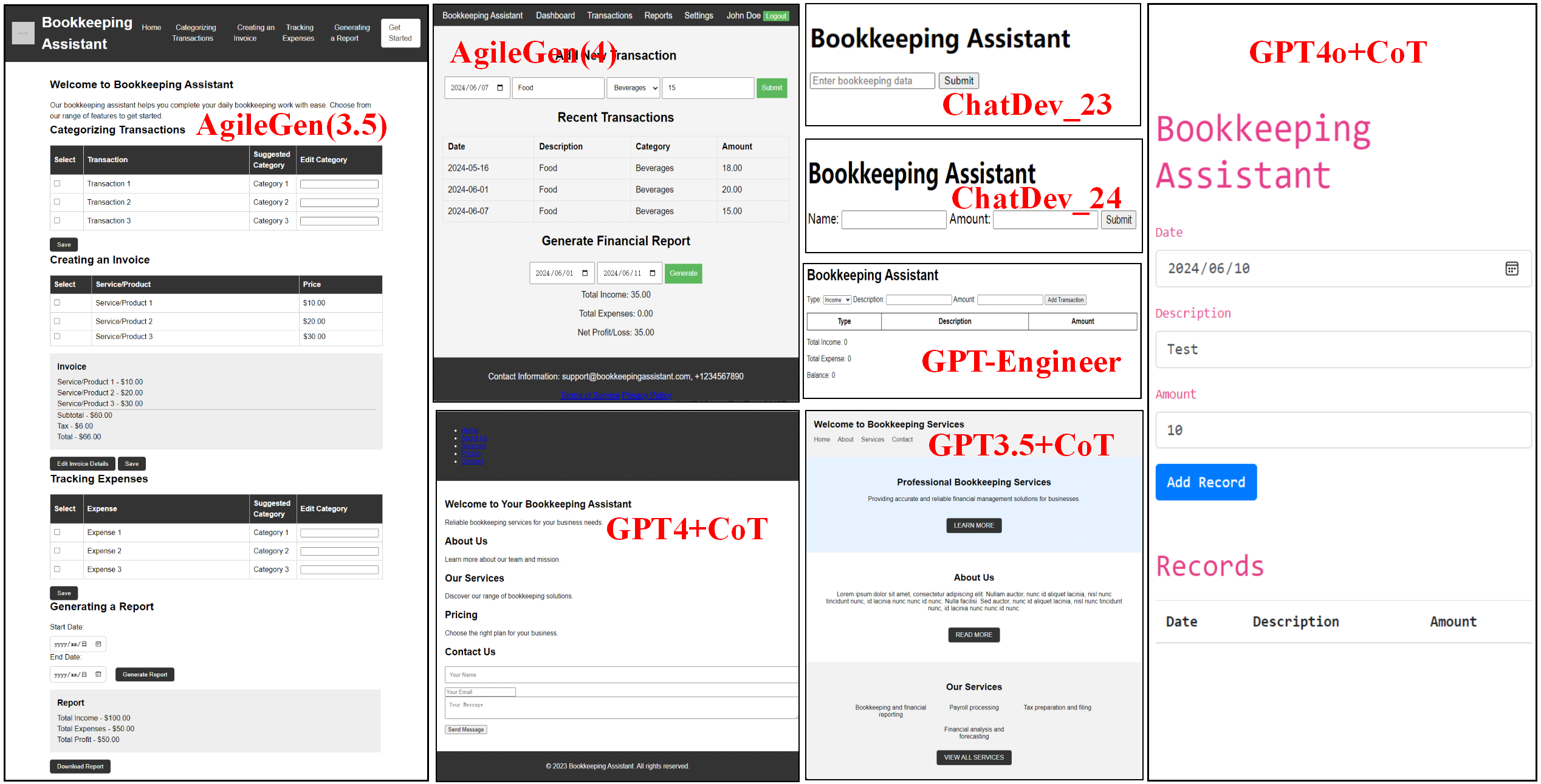}}
\caption{
Comparison of the generation effects of different methods on the real website application case ``Bookkeeping Assistant".}
\label{fig:methodcase}
\end{figure}

First, we present a real software case ``Bookkeeping Assistant,"~\footnote{https://github.com/HarrisClover/AgileGen-50projects50days-Experiment/tree/main/bookkeeping} in the ``50projects50days" dataset. Different methods generate a website application with the same task description, ``I need a bookkeeping assistant website." as input, and the effects are shown in Figure~\ref{fig:methodcase}. AgileGen (GPT3.5 and GPT4) and GPT-Engineer are capable of fulfilling the bookkeeping function. AgileGen(3.5) generates more features, while AgileGen(4) generates more concise features, completing the core functionalities with a more reasonable page layout. ChatDev, GPT3.5+CoT and GPT4+CoT generated pages that do not complete the bookkeeping function. Instead, they are combinations of common webpage modules (e.g., ``About US" and ``Contact US"). This may be due to many such modules in the large language models' training corpus. The pages generated by GPT4o+CoT use Bootstrap style sheets, but the functionality is incomplete and unable to record entries.

\begin{figure}[h]
\centerline{\includegraphics[width=\textwidth]{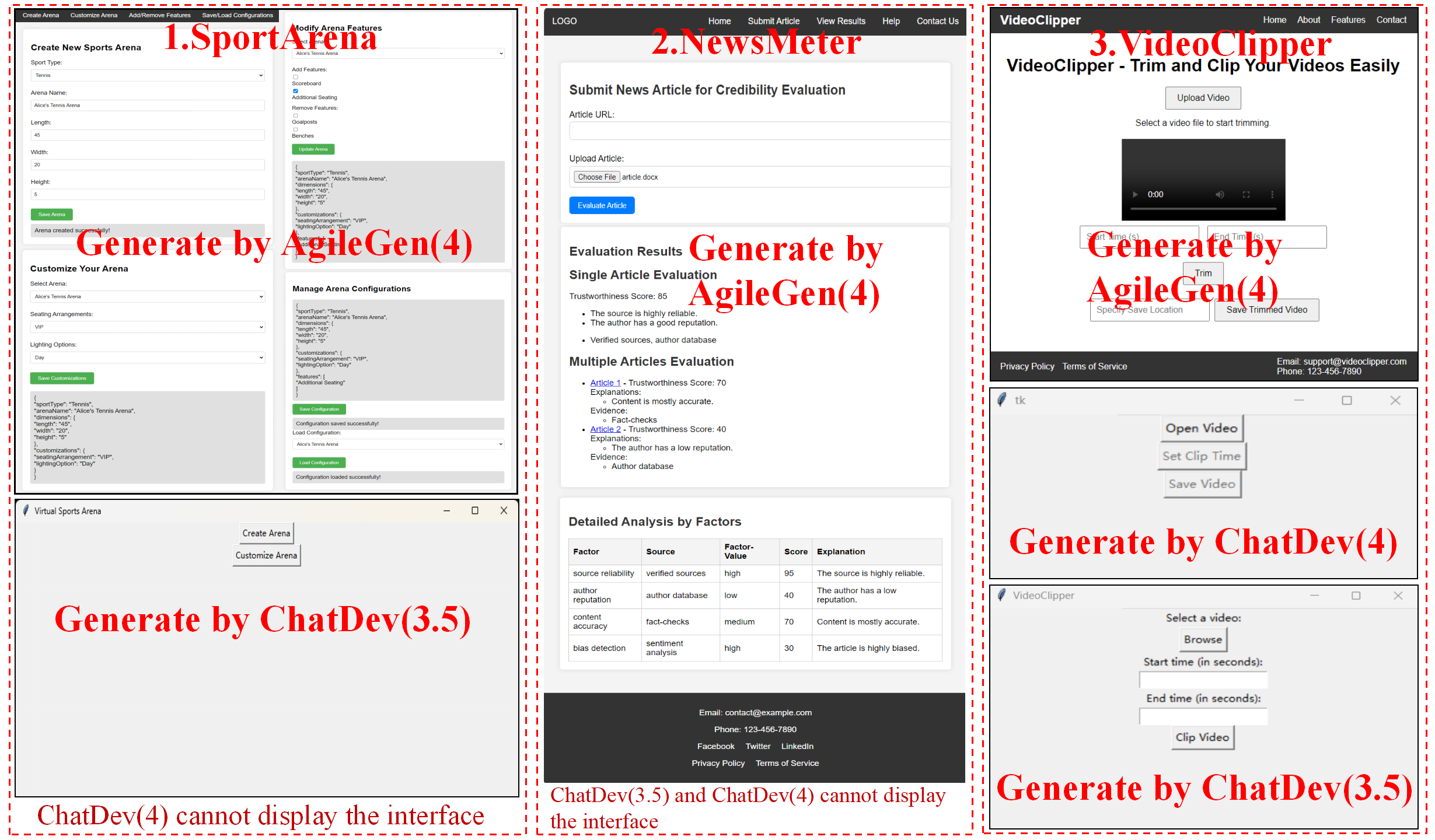}}
\caption{
Comparison of the generation effects of different methods on the cases ``SportArena",``NewsMeter" and ``VideoClipper" in ``SRDD" dataset.}
\label{fig:case_SRDD}
\end{figure}

In the "SRDD" dataset, Figure~\ref{fig:case_SRDD} shows the generation cases for three projects: "SportArena," "NewsMeter," and "VideoClipper." 
\begin{itemize}
    \item The task description for "SportArena"~\footnote{https://github.com/HarrisClover/AgileGen-SRDD-Experiment/tree/main/SportArena} is "Develop a user-friendly software application that allows users to create and customize virtual sports arenas." Our AgileGen method enables arena creation, feature customization, and configuration management, while ChatDev (GPT-4) generates an application without a GUI, and ChatDev (GPT-3.5) produces an interface with two non-responsive buttons labeled "Create Arena" and "Customize Arena."
    \item For "NewsMeter,"~\footnote{https://github.com/HarrisClover/AgileGen-SRDD-Experiment/tree/main/NewsMeter} the task is to "Evaluate the credibility of news articles by analyzing multiple factors and generating trustworthiness scores with explanations and evidence." Our method creates an application where users can upload articles and view evaluation results by clicking the "Evaluate Article" button. Unlike ChatDev, which fails to produce an effective interface, our approach simulates the analysis process, which we find reasonable. Evaluating news credibility is complex and may require integrating APIs and providing detailed information. By simulating these functional processes, our method provides a solid foundation for developing web applications with advanced features.
    \item For "VideoClipper"~\footnote{https://github.com/HarrisClover/AgileGen-SRDD-Experiment/tree/main/VideoClipper} is a software application that allows users to easily clip and trim videos. It provides an intuitive interface to select specific sections of the video and saves the trimmed video as a new file." All methods generate projects capable of video clipping, but our method demonstrates better interface design and interactivity.
\end{itemize}

\vspace{10pt}
\begin{mdframed}[backgroundcolor=white, roundcorner=5pt, leftmargin=1cm, rightmargin=1cm, innertopmargin=5pt, innerbottommargin=5pt]
\textbf{} \\
\textbf{Findings in RQ1}:
The experiments demonstrate that AgileGen surpasses all existing methods regarding the functional completeness of the generated software. The generated website applications achieved a high level of user satisfaction. Moreover, in the different case studies, our method is confirmed to fulfill the requirements and generate functions related to the requirements. This validates the effectiveness of our proposed method using Gherkin scenario design in the end-user requirements clarification.

\end{mdframed}

\subsection{\textbf{RQ2: What is the reason for the better performance of our framework?}}
\paragraph{\textbf{Motivation}}
The base model of the generative software development agent originates from the large language models GPT-3.5 or GPT-4. Given the inherent powerful performance of large language models, it is necessary to explore the comparative effects of our framework, AgileGen, against the baseline models. Additionally, we are interested in the effect of changes brought about by each part of our framework.

\paragraph{\textbf{Methodology}}
To explore RQ2, we will first compare our ablated methods on the "50projects50days" dataset using automatic and human evaluation metrics. Then, we will compare user satisfaction using the User Experience Questionnaire (UEQ). Finally, we will showcase the visual comparison of pages through case studies. We will introduce the exploration method for RQ2 according to a. Evaluation Metrics, b. Baselines, and c. Experimental Setup.

a. Evaluation Metrics. We use the automatic evaluation metrics CodeBLEU and Pass@k, human evaluation metrics Code Executability, and the User Experience Questionnaire (UEQ) to assess our ablation methods. Refer to section RQ1~\ref{subsection:RQ1} for detailed metrics descriptions.

b. Baselines. Each module of AgileGen will be ablated one by one until only the base model remains. The methods after ablation are described as follows:
\begin{itemize}

    \item \textbf{Our w/o Gherkin}: This method removes the entire scenario design component from AgileGen, as well as the consistency factor, automatic modification module, and design/code modification module from the rapid prototype design component shown in Figure~\ref{fig:scRapid}.
    After a simple prompt of ``Please expand the requirements," the expended end users' requirements are entered into the visualization module, where prompts within both the visualization and code generation modules have removed all Gherkin-related prompts.
    This method explores the impact of using the Gherkin language for user stories on code generation.
    \item \textbf{Our w/o Visual Design}: This method removes the visual design component from the rapid prototype design component, adjusting the code generation module to accommodate modifications.
    \item \textbf{Our w/o Consistency Factor}: This method removes the consistency factor module and the automatic modification module from the rapid prototype design component, used to explore the importance of Gherkin in ensuring the consistency of the generated software product with the end-users requirements.
    \item \textbf{Our w/o Anything}: This method removes all components from our method. It is a model without any extra prompting that directly inputs the user's task description into GPT3.5 or GPT4, subsequently generating software code. 
\end{itemize}

c. Experimental Setup. \textbf{Participants}: This experiment explores the impact of each module of our method on the generated software applications. Since this is a human evaluation of the software applications generated, we will again invite the 20 participants from RQ1.2 to evaluate our method's different ablation methods. For detailed information on the participant distribution, please refer to section~\ref{RQ1.2}. These participants have never had any conflicts of interest with us and have never used our method to develop software applications, ensuring their evaluations are unbiased.

\textbf{Study procedure}: The ablation experiment evaluation will be conducted simultaneously with the human evaluation experiment in RQ1.2. For details on session training and other aspects, please refer to section~\ref{RQ1.2}. Our methods with ablation will automatically generate web applications on the "50projects50days" dataset through scripts. The generated web applications will be placed into their corresponding folders and anonymized, retaining only the method ID. This ensures that participants will not know which method generated the web application they will be evaluated. Following the project allocation method from RQ1.2, participants will score the web applications included in each project. We will then summarize the results based on the method ID.

\paragraph{\textbf{Result Analysis}}
First, we will compare the ablated methods using the automatic evaluation metrics CodeBLEU and Pass@1 and the human evaluation metric Code Executability.
As shown in Table~\ref{tab:ablate}, we explore the contribution of each part of the framework to performance. \textbf{Our w/o Gherkin} shows a noticeable decrease in performance, indicates that user story formed by Gherkin language are very important in generating software code for meeting user requirements. \textbf{Our w/o Visual Design} method, although not significantly dropping in automatic evaluation metrics, does see a decline in manual evaluation scores, reflecting the significant impact of visual effects on user experience. The performance decrease in \textbf{Our w/o Consistency Factor} signifies the importance of Gherkin in ensuring the consistency of user requirements. 
The performance of \textbf{Our w/o Anything}, which includes only the base model, significantly declines, demonstrating that the superior performance of our method stems from the contributions of our framework's design rather than relying on the capabilities of large language models.

\begin{table}[t]
\caption{
Comparison of ablation study results.}
\begin{tabular}{llll}
\hline
\multicolumn{1}{c}{} & CodeBLEU & {Pass@1} & Code executability \\ \hline
\textbf{AgileGen (GPT3.5)} & \textbf{0.339} & \textbf{62.5\%} & \textbf{2.250} \\
Our w/o Gherkin & 0.143 (-0.196) & 20.6\% (-41.9\%) & 1.325 (-0.925) \\
Our w/o Visual Design & 0.292 (-0.0470) & 54.0\% (-8.5\%) & 1.800 (-0.450) \\
Our w/o Consistency Factor & 0.258 (-0.081) & 47.0\% (-15.5\%) & 1.550 (-0.700) \\ 
Our w/o Anything (GPT3.5) & 0.136 (-0.203) & 6.3\% (-56.2\%) & 1.175 (-1.075) \\
Our w/o Anything (GPT4) & 0.205 (-0.134) & 35.2\% (-27.3\%) & 1.525 (-0.725) \\ \hline
\end{tabular}
\label{tab:ablate}
\end{table}

\begin{figure}[h]
\centerline{\includegraphics[width=0.8\textwidth]{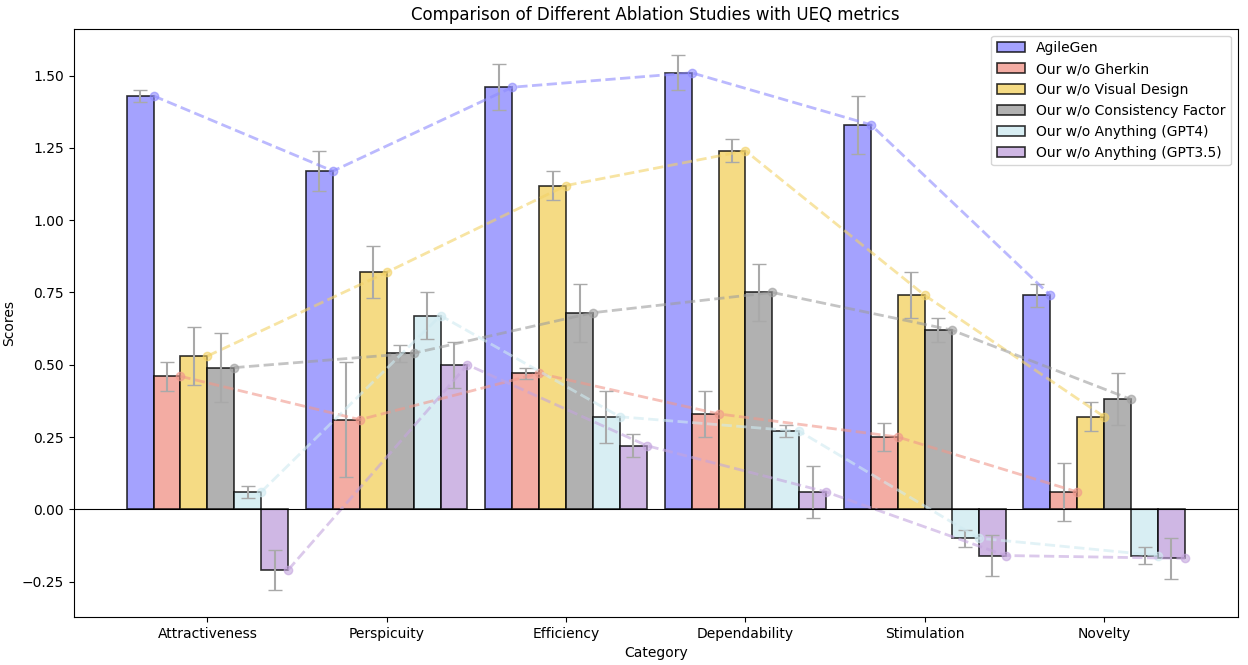}}
\caption{Comparison experiment results on UEQ metrics. The results of our method are all based on GPT-3.5.}
\label{fig:UEQ_ab}
\end{figure}

Next, we used the User Experience Questionnaire (UEQ) to compare our method with our ablation methods, as shown in Figure~\ref{fig:UEQ_ab}. Ablating the scenario design module (Our w/o Gherkin) resulted in a significant decline compared to AgileGen, especially in the "Perspicuity" metric, which is because, without acceptance criteria, it is difficult for large language models to define functional boundaries, often leading to generate the ambiguous function blocks. Ablating the visual design module (Our w/o Visual Design) caused a notable decrease in the "Attractiveness" metric, while other metrics experienced minor declines. Web applications without visual design often appear cluttered and lack stylistic content. The slight decline in functionality metrics may be due to a few features in our "50projects50days" dataset relying on visual effects. Metrics for Our w/o Consistency Factor also showed a notable decline, demonstrating Gherkin's significant role in generating applications that align with user requirements. The minimal decline in the "Novelty" metric suggests that Gherkin introduces surprising features by supplementing user requirements with acceptance criteria. Lastly, removing all modules (Our w/o Anything) resulted in significant decreases across all metrics. The notable declines in "Dependability," "Stimulation," and "Novelty" indicate that directly providing large language models with requirements lacking acceptance criteria fails to meet user needs.
In summary, our method uses Gherkin to supplement unclear user requirements with well-defined acceptance criteria, guiding the generation of applications that align with user requirements. Additionally, our framework enhances visual design, improving the end-user experience.

\begin{figure}[tbp]
\centerline{\includegraphics[width=\textwidth]{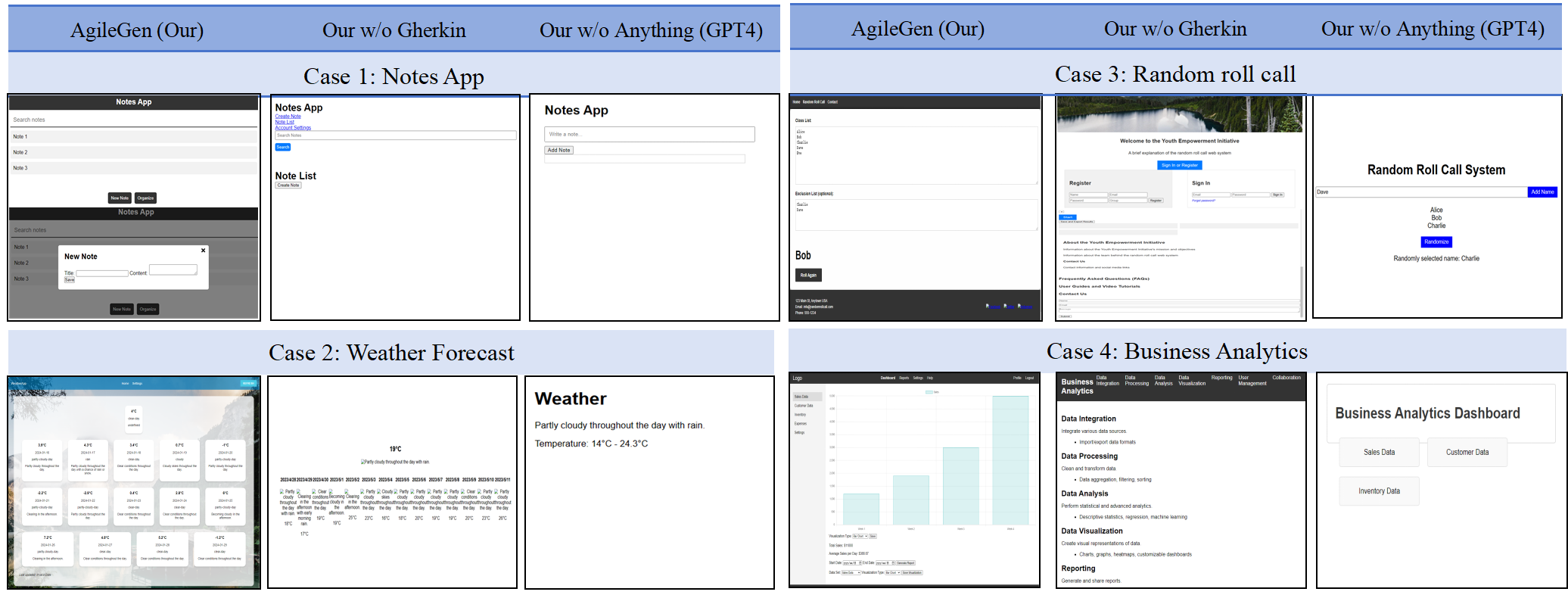}}
\caption{Case study comparison diagram. The case studies 1, 2, and 3 are from the "50projects50days" dataset, while case study 4 is from the "SRDD" dataset.}
\label{fig:example}
\end{figure}
Finally, we compare the interfaces of software through the case studies. In Figure~\ref{fig:example}, from left to right, are "AgileGen," "Our w/o Gherkin," and "Our w/o Anything (GPT4)." We provide detailed examples below:
\begin{itemize}
    \item \textbf{Case 1:Notes App.}
Our framework is designed to provide an efficient and clean software interface, including the main functions of browsing notes and adding and editing them. However, if the Gherkin is removed (Our w/o Gherkin), the software applications generated lacks functionality and does not respond to clicks. In Our w/o Anything (GPT4), The software function is imperfect due to the lack of acceptance criteria, with only the simple "add note" function available.
    \item \textbf{Case 2: Weather Forecast.}
Our framework creates visually appealing software showcasing the city's current and future weather conditions. In Our w/o Gherkin, the software's functionality is adequate, but the page aesthetics suffer. In Our w/o Anything (GPT4) can only provide information about daily weather conditions. This is functionally inadequate, and its page aesthetics are relatively poor.
    \item \textbf{Case 3: Random roll call.}
Our framework fulfills end-user requirements for the random roll call. It includes adding, deleting, and modifying the class name and exclusion lists for leave requests. However, in Our w/o Gherkin, the generated pages may have redundant parts (Registration, Login, Contact Us) because common modules like "registration" and "login" often appear in the training data of large language models. Our w/o Anything (GPT4) generates limited functionality, requiring manual name additions without a delete option.
    \item \textbf{Case 4: Business Analytics}
    Our framework designed a feature-rich web application that accomplishes business analytics data visualization. The generated project allows switching between different data visualization types and includes features for exporting reports and saving visualizations. Although Our w/o Gherkin retains a clean page appearance, its functional blocks are listed without implementation. Our w/o Anything (GPT4) fails to perform basic functions, with non-responsive buttons and an unattractive page layout.
\end{itemize}

\vspace{10pt}
\begin{mdframed}[backgroundcolor=white, roundcorner=5pt, leftmargin=1cm, rightmargin=1cm, innertopmargin=5pt, innerbottommargin=5pt]
\textbf{} \\
\textbf{Findings in RQ2}:
Our AgileGen uses Gherkin to bridge the gap between end-user requirements and the technical implementation by the large language models. Gherkin supplements unclear user requirements with well-defined acceptance criteria in the form of user stories, guiding large language models in defining the boundaries of the generated code functionalities. The ablation studies verifying that each module within our AgileGen positively impacts performance. The comparison with the Our w/o Anything shows that the high performance of our framework stems from the contributions of our framework's design.
\end{mdframed}

\subsection{\textbf{RQ3}: What is the human-computer interaction experience for end-users during the AgileGen  process?}
\label{Sec:RQ3}
Previous RQs evaluated our method's performance in generating software applications without human interaction for fairness. However, our method ensures semantic consistency between user requirements and agent generation and supports end-user participation in software development. It enables collaborative development using the behavior-driven testing language Gherkin to clarify user stories through user-agent interactions. Agile methodologies inspire this iterative approach and incorporates user feedback to enhance project development. Thus, it's essential to explore our framework's effectiveness in interactive experiences and the impact of human interaction on generating software applications.
\subsubsection{RQ3.1: How effective is our method regarding human-computer interaction experience metrics?}
\paragraph{\textbf{Motivation}}
The human-computer interaction experience of end-users is also crucial for generative software development agents. Therefore, it is necessary to explore the effectiveness of our framework in terms of interaction experience.
\paragraph{\textbf{Methodology}}
To explore RQ3.1, we will introduce the following aspects: a. Human Evaluation Metrics, b. Baselines, and c. Experimental Setup.

a. Human Evaluation Metrics: We use the Likert Scale~\cite{DBLP:conf/chi/WuTC22[61]} to evaluate our method's human-computer interaction (HCI) experience. HCI is crucial for measuring the AgileGen, as users interact with it while generating website applications. We used the User Interaction Experience Scale by Wu et al. ~\cite{DBLP:conf/chi/WuTC22[61]}. to assess HCI, which was scored on a seven-point Likert scale. This scale includes questions on "Match Goal," "Think Through," "Transparent," "Controllable," and "Collaborative."

b. Baselines. We will test our method on different base models: one based on GPT-3.5, referred to as Our (3.5), and another based on GPT-4, referred to as Our (4). These two versions have identical implementation details, and we can switch between different models by modifying the `model' parameter in the tool configuration.

c. Experimental Setup. \textbf{Participants}: 
To conduct a human evaluation of our method's human-computer interaction experience, we recruited 20 individuals from various industries, different from the participants in RQ1.2. We designed a detailed email invitation explaining the study's purpose and tasks and offering a small financial compensation of \$20 per person. Twelve of these participants were male, and eight were female. 
45\% came from various professions, including Consulting Media, Design Industry, Education, Finance, Medical Treatment, Service, and Online Retailers. 55\% were students from other institutions: 6 majoring in software engineering, 1 in computer science and technology, 2 in new media and communication, and 2 in network engineering. Regarding age distribution, 55\% were aged 18\~{}24, 35\% were aged 25\~{}34, and 10\% were aged 35\~{}44. Among the 20 participants, 5 had only basic computer operation knowledge, 7 were novice programmers (0\~{}1 year of experience), 4 were beginners (1\~{}3 years of experience), and 4 were experienced programmers (more than three years of experience). These participants had never been involved in any stage of our development and had no overlap with the participants from RQ1.2, ensuring their evaluations were objective and unbiased.

\textbf{Study procedure}: To ensure participants understand the Likert Scale scoring rules and our method's usage and interaction process, we organized an online training session for the 20 participants. We selected one web application each from the "50projects50days" and "SRDD" datasets not chosen for evaluation to show our interaction process, including user decision-making and iterative acceptance and recommendation interactions.
During the training session, we simulated the scoring operation using the Likert Scale. The session included a Q\&A segment to confirm that participants had no doubts. After the training, each participant was randomly assigned to test six projects from the two datasets using our method. Participants were allowed to modify and improve features according to their needs. After testing each project, participants were asked to rate it using the Likert Scale scoring form. Finally, we interviewed the participants to gather their feedback on our method.
\paragraph{Result Analysis}

\begin{figure}[t]
\centerline{\includegraphics[width=\textwidth]{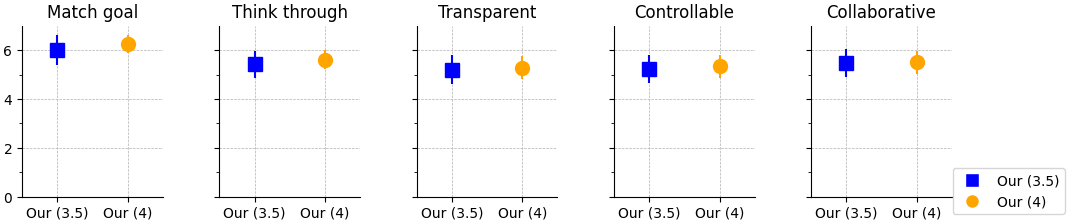}}
\caption{Comparison of human-computer interaction experience results. Each coordinate represents a dimension of interaction. The square markers represent the results of our method based on GPT-3.5 (Our (3.5)), while the circular markers represent the evaluation results of our method based on GPT-4 (Our (4)).}
\label{fig:interaction}
\end{figure}

As shown in Figure~\ref{fig:interaction}, AgileGen performs well in human-computer interaction, scoring above average on all five metrics, with high stability and a notable "Match goal" score based on GPT-4. Participants praised its controllability, noting they could modify the web application as needed and add or remove features after the initial generation. Many found using natural language for web project development enjoyable, allowing them to focus on planning features rather than implementation details. Most participants engaged in about three interaction modifications (here, interaction refers to scenarios decision-making and acceptance \& recommendation decision-making, not the number of button clicks) per project in the experiment. Interestingly, some gave a high "Match goal" score for the "09 Sound Board" project, which had a 0\% task completeness score in the automatic evaluation due to the lack of multimedia audio. One participant's simple request, "I want to be able to click on a card and read it automatically." led to iterative modifications using the Web Speech API's "speechSynthesis" for speech synthesis. In summary, our method is "transparent" and "controllable," allowing easy requirement modifications and improving the "Match goal" metric through iterative processes. It engages end-users in decision-making, making it more "collaborative" and beneficial for those lacking software development expertise.

\subsubsection{RQ3.2: How does the human-agent collaboration process affect the generation of software applications?}

\paragraph{\textbf{Motivation}}
The above experiments demonstrate that users have a good experience when interacting with our method in human-agent interactions. Although the "Match goal" metric can reflect the impact of human-agent collaboration on the generated results, it is not very intuitive. Exploring the impact of human-agent interaction on the generation of software applications through case studies will provide a more direct view of the changes and impacts during interactions, showcasing user-decision details.

\paragraph{\textbf{Methodology}}
We find three projects from the participant evaluations that performed poorly in automatic generation but showed significant improvement after the human-agent collaboration process. We will showcase these three projects. Finally, we demonstrate the robustness of our method using the task description from~\cite{DBLP:journals/corr/abs-2307-07924[1]2}: "design a basic Gomoku game." We have prepared a replication package~\footnote{https://github.com/HarrisClover/AgileGen-Human-Agent-Interaction-Experiment} for public validation of the experiment.

\paragraph{\textbf{Result Analysis}}
\begin{figure}[th]
\centerline{\includegraphics[width=0.9\textwidth]{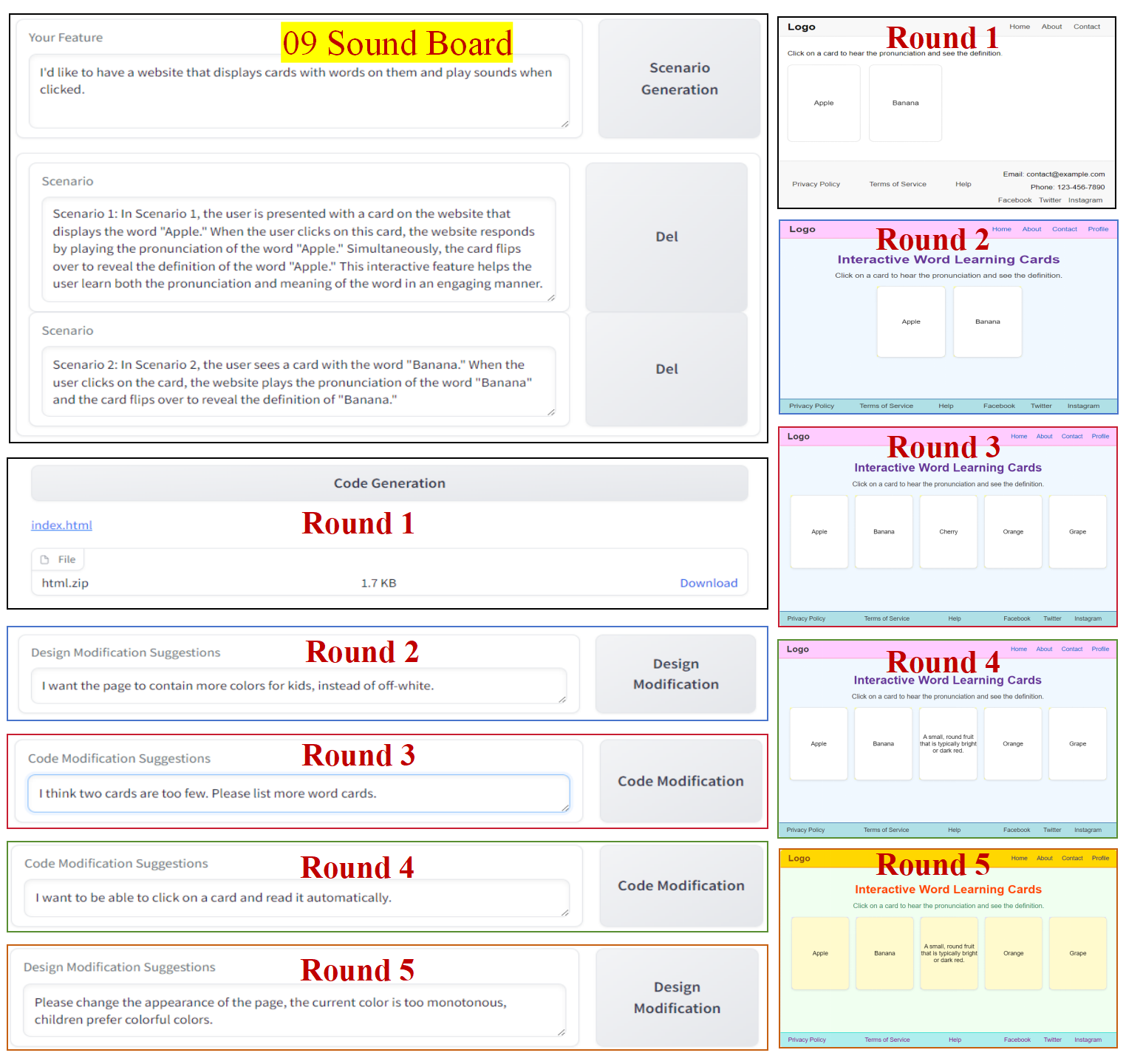}}
\caption{The interaction showcases for the "09 Sound Board" project includes AgileGen's interaction system on the left and the generated end-user web application interface on the right. The participant made five iterative modifications to the project.}
\label{fig:interaction_case_1}
\end{figure}

The "09 Sound Board" project~\footnote{https://github.com/HarrisClover/AgileGen-Human-Agent-Interaction-Experiment/tree/main/09\%20Sound\%20Board} showcases the user decision-making process, as illustrated in Figure~\ref{fig:interaction_case_1}. The initial task description is: "I'd like a website that displays cards with words and plays sounds when clicked." Due to the need for multimedia audio, no methods complete the sound playback functionality during automatic generation. We recorded a participant's entire testing process, consisting of five rounds of iterative modifications. Initially, two scenarios are provided, and the participant makes no changes. The "Round 1" result shows a predominantly gray and white theme. The participant then suggests a visual modification: "I want more colors for kids, instead of off-white," leading to a colorful page in "Round 2."  Next, the participant requests more word cards: "I think two cards are too few. Please list more word cards." In "Round 3," the number of cards increases to five, maintaining consistent themes. In the fourth iteration, the participant asks for an automatic reading feature: "I want to be able to click on a card and read it automatically," resulting in a modification using the Web Speech API to read the card. Finally, the participant suggests another visual change, preferring bright colors. In "Round 5," the page is adjusted to use brighter orange tones, including card colors. This interaction demonstrates that participants can freely modify functionality and visual effects using natural language. Moreover, the human-agent interaction process helps complete the generation of the software application's functionalities.

\begin{figure}[t]
\centerline{\includegraphics[width=0.9\textwidth]{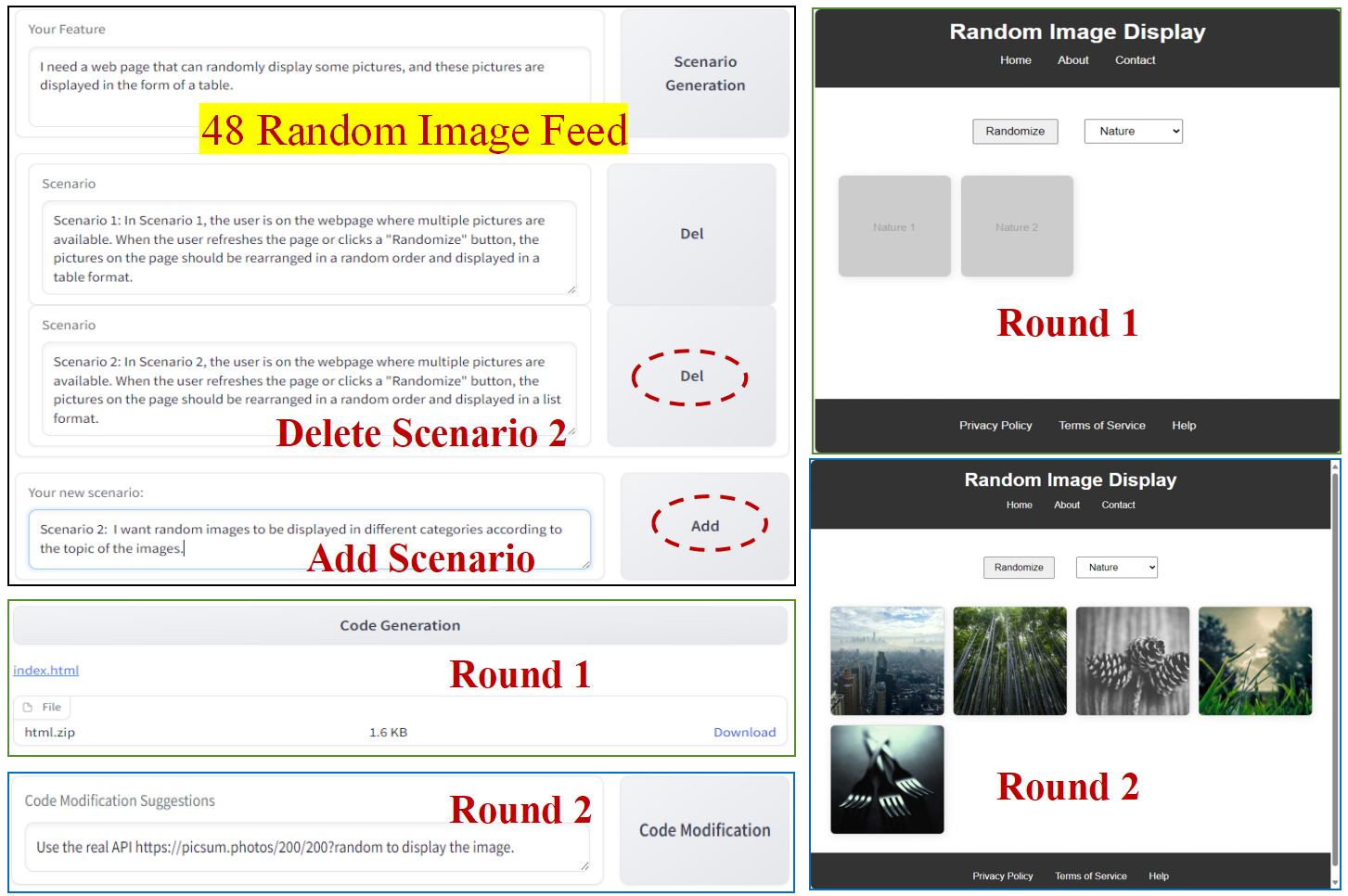}}
\caption{The interaction showcases for the "48 Random Image Feed" project includes AgileGen's interaction system on the left and the generated end-user web application interface on the right. The participant made scenarios decision-making and went through 2 rounds of iterative modifications.}
\label{fig:interaction_case_2}
\end{figure}

Next, the "48 Random Image Feed" project~\footnote{https://github.com/HarrisClover/AgileGen-Human-Agent-Interaction-Experiment/tree/main/48\%20Random\%20Image\%20Feed} is chosen as the second example, illustrated in Figure~\ref{fig:interaction_case_2}. The initial task description is: "I need a web page that can randomly display some pictures, and these pictures are displayed in the form of a table." This project focuses on image arrangement and display. During participant interaction, an interesting process unfolds. The participant deletes the scenarios they deem unnecessary and adds their scenario description: "Scenario 2: I want random images to be displayed in different categories according to the topic of the images." The "Round 1" result includes the requested categorization, but the images are placeholders. Then, the participant provides an API link as part of the modification description to enhance functionality: "Use the real API https://picsum.photos/200/200?random to display the image." The "Round 2" result shows the web application using this external API to display images. This example demonstrates our method's ability to integrate custom APIs to achieve more complex functionalities.

\begin{figure}[th]
\centerline{\includegraphics[width=\textwidth]{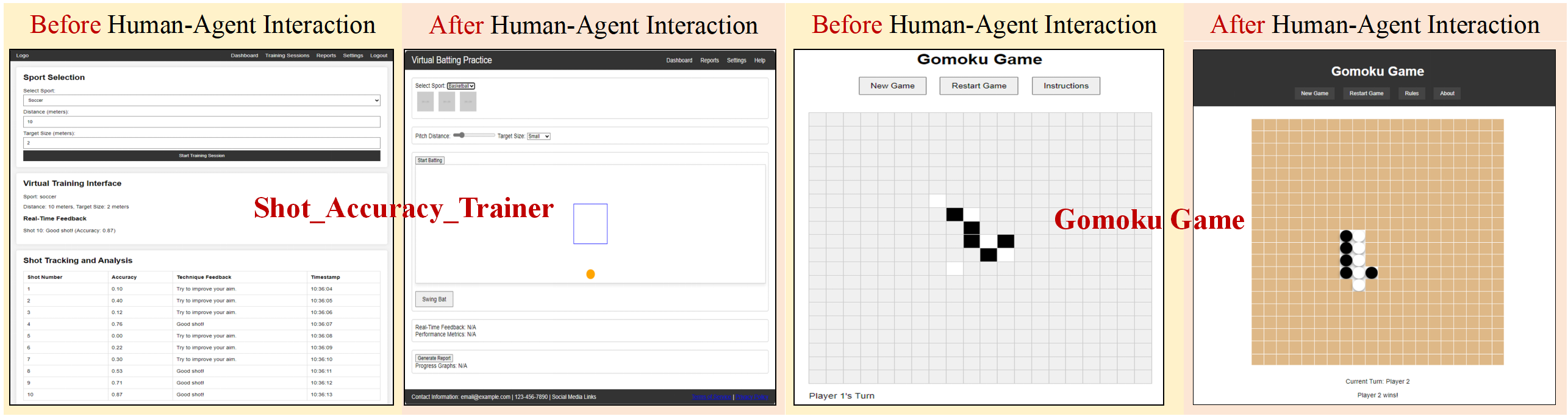}}
\caption{The results are shown before and after interaction for the "Shot\_Accuracy\_Trainer" and "Gomoku Game" projects. On the left is the state before human interaction, and on the right is after the interaction.}
\label{fig:interaction_case_3}
\end{figure}

Finally, we compare the web application before and after human interaction, as shown in Figure~\ref{fig:interaction_case_3}. 
In the "Shot\_Accuracy\_Trainer" project~\footnote{https://github.com/HarrisClover/AgileGen-Human-Agent-Interaction-Experiment/tree/main/Shot\_Accuracy\_Trainer}, the initial generation tends to create a record-keeping website, overlooking that it is a basketball shooting game. User interaction clarifies (The Code Modification suggested: "Virtual training is a shooting game. I can click the button after the force has a small ball that can shoot.") that it is a game and generates the corresponding shooting functionalities.
In the "Gomoku Game" project~\footnote{https://github.com/HarrisClover/AgileGen-Human-Agent-Interaction-Experiment/tree/main/Gomoku}, the initial generation has already completed the rules and functionality of the Gomoku game. After the interaction, the board and pieces appear more aesthetically pleasing.

\vspace{10pt}
\begin{mdframed}[backgroundcolor=white, roundcorner=5pt, leftmargin=1cm, rightmargin=1cm, innertopmargin=5pt, innerbottommargin=5pt]
\textbf{} \\
\textbf{Findings in RQ3}: 
Our AgileGen has established a user-friendly interaction interface for end-users. The human evaluations have shown that our method can better achieve its objectives, enhancing the sense of participation and collaboration for end-users during the system operation process. The case studies clearly and intuitively demonstrate the participants' interaction processes and the changes in the generated web applications. In summary, our method can generate high-quality software code and offer a superior interaction experience.
\end{mdframed}

\section{Discussion}
In this section, we discuss the introduction of external tools, the hidden costs of user decisions, and the scalability and generalizability of our framework, based on some error analysis.

\textbf{External tools with Discussion:}
We discuss the issue of external tools on two levels: the Potential benefits of integrating external tools into Our method and the benefit of introducing external tools to generate software applications.

\textit{1) Potential Benefits of Integrating External Tools into Our Method:} 
Our method integrates the SQLite3 external database tool in the memory pool to save human decision experiences. We chose SQLite3 because it is an open-source, lightweight, embedded relational database that requires no server configuration and supports most SQL syntax, enabling easy access to stored experiences. If we need higher concurrency or more complex queries in the future, SQLite3 can easily integrate with server-side database management tools.
Our approach is based on behavior-driven development (BDD), where the scenario design module includes user-defined Gherkin scenarios as test cases. These scenarios can be tested using the external Cucumber tool, which would provide feedback to improve code quality. Since Gherkin ensures alignment between code and business logic, integrating Cucumber can help create software that meets user expectations, reducing manual iterations.
For example, in the "09 Sound Board" project, the initial scenario specifies, "Then the website plays the pronunciation of the word `Apple.'" However, in "Round 1," this functionality is not generated. By introducing Cucumber, the tool could automatically test and correct such functionality, minimizing manual corrections. This will be a key focus of our future research.

\textit{2) Benefits of Introducing External Tools to Our Generated Software Applications:}
During software applications generation, large language models (e.g., GPT-4) have hallucination problems, resulting in generated real multimedia API failure. To maintain the display effect of web application interfaces, we typically replace multimedia resources with placeholders during automatic generation. However, some projects rely on images to present their visual effects. Placeholder images are often gray and provide a poor visual experience, which can result in participants giving lower visual scores even if the functionality is complete.
For example, the project "ID50 Catch The Insect" should involve clicking on insect images to catch them, but the game theme can be confusing without the insect images. Similarly, the "ID48 Random Image Feed" project requires multimedia resources for display, but automatically generated resource links often do not meet the requirements. Our method allows for specifying certain resource APIs through human-agent collaboration, as shown in Figure~\ref{fig:interaction_case_2}. This interaction process made us realize that incorporating external API tools can enable our generated end-user applications to achieve more complex functionalities.
In the "SRDD" dataset, for instance, the "NewsMeter" project aims to complete a news credibility analysis function, as shown in Figure~\ref{fig:case_SRDD}. Our method generates a reasonable simulation process, including data display interfaces in the code. If a suitable news analysis API is introduced, it could easily result in a genuine news analysis website system. Of course, we need to consider the reliable sources and stable references of APIs and the different response schemas of each API. Completing the functionality of an API with complex input and output data may require further restricting the generation format of software applications.
In summary, the introduction of external API tools would enhance the scope of applications generated by our method, and we will consider this in our future work.

\textbf{Hidden Costs of User Decisions}:
In our method, users can participate in decision-making steps, including "end-user requirements decision-making," "scenario decision-making," and "acceptance \& recommendation decision-making." Therefore, we discuss these in three parts.

\textit{1) ``End-User Requirements Decision-making":} This is when users propose the initial requirements for the software applications in our system. Our method accommodates requirements that are simple natural language descriptions, as shown in Table~\ref{tab:10data}, rather than complex requirement specifications. This reduces the domain knowledge and learning costs required from end-users during the requirements decision-making.

\textit{2) ``Scenario Decision-making":} This involves users adding, deleting, or modifying the generated Gherkin scenarios in our system to clarify their requirements further. 
We designed a memory pool module to reduce the number of decision iterations caused by poor scenario quality. This module accumulates previous user decision experiences and recommends similar scenarios for future similar projects.
However, to ensure fairness in the experimental setup, we did not provide any human experience before testing all projects, so the impact of the memory pool can only be measured later. Our design mechanism accumulates human experiences over time, reducing user decision costs as more tests are conducted. 
In the future, we plan to deploy our method on a public testing platform to collect more human experiences over time, demonstrating the process of continuous enhancement through human-in-the-loop interactions. Additionally, we built an interaction bridge to reduce the learning cost for users when adding and modifying Gherkin scenarios, allowing users to add and modify functionalities using natural language seamlessly.

\textit{3) ``Acceptance \& Recommendation Decision-making":} This allows users to review the software application and iteratively recommend modifications in our system. To reduce the learning cost for users when executing applications during acceptance, we set up a "Code Execution Link" in the system. Users can click the link easily to test the web application interface. To reduce the number of iterations for user recommendation decisions, we designed a consistency factor that automatically modifies the code to improve alignment with user requirements. Lastly, we can further reduce the number of manual iterative recommendation modifications in the future. Since our method is based on behavior-driven development (BDD), we can easily introduce the external tool Cucumber for automated testing of the generated code and targeted modifications. This approach will significantly enhance the consistency between implementation code and business logic, reducing user decision costs during "acceptance \& recommendation decision-making."

\textbf{Scalability and Generalizability Discussion:}
\textit{1) In terms of scalability:} First, our method is based on Gherkin scenario test cases, which can easily be extended to include external Cucumber testing tools. This targeted modification of the generated code would significantly improve code quality. Second, when facing future high concurrency and complex queries in the human experience memory pool, the SQLite3 database can be expanded to MySQL or other server-side data management systems. Finally, our system harnesses the capabilities of large language models but is not limited to a specific model. Thus, it can be easily extended to various models, demonstrating high scalability.

\textit{2) In terms of generalizability:} First, our framework can be applied to any software development project. By adopting behavior-driven development (BDD) based on agile principles, we ensure the consistency of generated code with business logic. This innovative behavior-driven code generation approach is not confined to web application development. If we want our framework to support mobile application development, we can change the visual design components and the programming language of the code generation components. Finally, our method has received high user satisfaction and interaction experience ratings from participants in various industries, validating its generalizability in generating software applications. Thus, our method demonstrates high generalizability in its framework design and final generated software application.

\section{Related Work}
This paper introduces AgileGen, an AI agent designed for generative software development based on Agile methodologies and human-AI collaboration. AgileGen enables both end-users and AI to focus on the tasks they excel at, working together to effectively complete the development process. The related works are as follows:
\subsection{Automated code generation}
In recent years, automated code generation tasks have been achieved using deep learning to generate code automatically~\cite{DBLP:conf/sigsoft/ShenZDGZL22[3],DBLP:conf/kbse/WangLZLLWC22[4],DBLP:conf/acl/WangWWMLZLWJL22[11]2,DBLP:conf/kbse/0001YN22[12]2,DBLP:conf/icse/YandrapallySTM23[13]2,DBLP:conf/sigsoft/BoyalakuntlaCC022[14]2}. 
Automated code generation tasks are divided into tree-based and token-based methods~\cite{DBLP:conf/sigsoft/ShenZDGZL22[3]}. Tree-based methods generate code by producing tree structures or Abstract Syntax Trees (AST) ~\cite{DBLP:conf/kbse/0001YN22[12]2,DBLP:conf/acl/DongL16[15]2,DBLP:conf/emnlp/YinN18[16]2}. Meanwhile, token-based methods~\cite{DBLP:conf/kbse/WangLZLLWC22[4],DBLP:conf/acl/WangWWMLZLWJL22[11]2,DBLP:conf/icse/YandrapallySTM23[13]2,DBLP:conf/sigsoft/BoyalakuntlaCC022[14]2} treat the input and output code as sequences of tokens. 
There is also a type of template-based code generation. \cite{10043059} is a tool for the automatic generation of web software and API code based on templates.
With the software of large pre-trained models in code generation tasks, token-based methods have shown remarkable results. CodeT5~\cite{DBLP:journals/corr/abs-2107-03374[21]} introduced identifier-aware pre-training tasks based on T5. CodeGPT~\cite{DBLP:conf/nips/LuGRHSBCDJTLZSZ21[17]2}, CodeBERT~\cite{DBLP:conf/emnlp/FengGTDFGS0LJZ20[20]}, and CodeGeeX~\cite{DBLP:journals/corr/abs-2303-17780[5]} utilize various programming languages for pre-training. CodeX~\cite{DBLP:conf/emnlp/0034WJH21[19]} is a GPT-based model that underwent pre-training on a large scale of GitHub code and launched a productivity version named GitHub Copilot~\cite{Github_copilot[18]2}. These powerful pre-trained models can generate high-quality code snippets during software development, enhancing developers' productivity through pair programming. This lays the foundation for automatically generative software development~\cite{DBLP:journals/corr/abs-2009-10297[10]2}. However, these large pre-trained code generation models mainly focus on generating high-quality code snippets within a single file.

\subsection{Generative Software Development}
Generative software development has long been a focus of attention in industry and academia. Paolone et al.~\cite{DBLP:journals/computers/PaoloneMPF20[1]} proposed a model-driven architecture (MDA)-based approach using Unified Modeling Language (UML) and xGenerator to generate website applications. Boyalakuntla et al.~\cite{DBLP:conf/sigsoft/BoyalakuntlaCC022[25]} aimed to reduce smartphone battery consumption by creating a meta-model and generating location-based software. Bernaschina et al.~\cite{DBLP:conf/icse/BernaschinaCF17[30]} developed a model-driven development tool that automatically generates rapid prototypes of web and mobile software from Interaction Flow Modeling Language (IFML) specifications. 

Large language models such as ChatGPT~\cite{DBLP:journals/corr/abs-2302-04023[11]} have been widely applied in software development and demonstrated potential. Microsoft researchers Bubeck et al.~\cite{DBLP:journals/corr/abs-2303-12712[10]} guide GPT-4 to the development of complex HTML games. Some Python project generators based on GPT have been released on Huggingface~\cite{Python_Snippet_Generator[31]}, generating function architecture based on brief requirements. AutoGPT~\cite{AutoGPT[32]} has been explored for generating HTML projects without manual prompting. Intelligent tools can help developers be more productive, but they need precise models and expert knowledge to work well. Similar projects include DemoGPT\footnote{https://github.com/melih-unsal/DemoGPT}, which utilizes the Streamlit visualization tool and can rapidly transform a single prompt into an AI software with the prompted functionality. This fully autonomous AI operates without human intervention in its internal processes and is prone to issues such as process freezing. 


Recently, the concept of multi-agent collaboration has been introduced. Generative Agents~\cite{10.1145/3586183.3606763} simulate human behavior by allowing agents to interact collaboratively through dialogue. CodeAgent~\cite{tang2024collaborative} conducts code reviews through the collaboration of multiple agents led by a QA-checker. Inspired by this collaborative interaction among agents, MetaGPT~\cite{DBLP:journals/corr/abs-2308-00352[2]2} utilizes standard operating procedures to coordinate a multi-agent system based on large language models, with different agents simulating different software development roles. Subsequently, MetaGPT proposed a solution for multi-agent collaboration in data science applications~\cite{hong2024data}. ChatDev first version~\cite{DBLP:journals/corr/abs-2307-07924[1]2} designs a dialogue chain-guided, multi-agent collaborative software development agent. Based on the original prototype of ChatDev, two versions were updated. The ChatDev second version~\cite{qian2023experiential} adopts experiential co-learning, collecting shortcut-oriented experiences from their historical trajectory for use in software development. The ChatDev third version~\cite{qian2024iterative} refines these historical experiences through iteration before using them in software development. The fourth version of ChatDev~\cite{qian2024scaling} uses a directed acyclic graph (DAG) to organize agents. Employing topological sorting simplifies agent interaction reasoning and enhances collaboration among agents. These agents follow the waterfall model, characterized by heavy documentation and process emphasis. However, they offer lower levels of human participation and control, and the hallucination errors of each agent can cascade, making it challenging to generate software products that meet user requirements.

Some questioning-based agents have been proposed to mitigate the accumulation of hallucination errors. These agents present problems that are difficult for them to solve back to humans in the way of questions, allowing humans to clarify or modify them. GPT-Engineer~\cite{Gpt-engineer[4]2} is a method for automated software generation using GPT-4, which clarifies requirements with users through questions before generating code, including issues like database requirements and API usage. Similarly, gpt-pilot~\footnote{https://github.com/Pythagora-io/gpt-pilot} generates software code after asking developers to clarify various domain-specific questions. Data Interpreter~\cite{hong2024data} uses code to solve data science-centered analytical problems. In its dynamic plan management module, two strategies are designed: self-debugging and human editing, where tasks that fail multiple self-debugging attempts are modified through human editing. These methods aim to assist professional developers who usually require domain knowledge. End-users with business requirements are often limited by domain knowledge and find it challenging to describe requirements from a development perspective. Our approach enables end-users and agents to collaborate effectively, each handling what they excel at while jointly completing software development.
 
\subsection{Requirement Generation}
Unclear requirements are a primary difficulty in generative software development, as end-users often need more domain knowledge to articulate their needs precisely. The requirements are generally expressed in a natural language of one to two sentences. The goal of the requirement generation task is to produce clear requirements. Existing research on automated requirement generation primarily focuses on converting (semi-)structured models into natural language requirement specifications with specific syntactic patterns~\cite{6836450[19]2,DBLP:journals/re/SouagMSC18[20]2,DBLP:journals/infsof/NicolasA09[21]2}. Few studies pay attention to generating requirement specifications from keywords provided by requirement analysts~\cite{math11020332[22]2}. They fine-tune UniLM by associating keyword knowledge and reinforcing relevancy, integrating a copy mechanism to meet the hard constraints of keywords, and employing grammar-constrained decoding to satisfy grammatical requirements. However, end-users struggle to possess the knowledge required to construct precise models or describe professional keywords.

This paper introduces an agile-based AI agent for generative app development through human-AI teamwork (AgileGen). It generates acceptance criteria presented in the Gherkin language from the incomplete requirements of end-users. The Gherkin is then translated into scenarios described in natural language for users to make decisions, thereby clarifying end-user requirements. The code generation component produces interface designs and consistency factors based on scenarios that have been decided upon. The consistency factors guide code generation to ensure the app product meets the end-user's requirements. The final, workable app product is presented to the end-users for acceptance and recommendations, and then the next iteration cycle is entered. Our AgileGen is designed for end-users who can easily generate satisfactory software without the need for domain knowledge.

\section{Conclusion}
This paper constructs a bridge for generative app development based on Agile methodologies between end-users and the Agent, enabling them to handle tasks in which they are proficient in teamwork. This approach avoids the error propagation caused by top-down waterfall model thinking and reduces the professional need for domain knowledge in clarifying requirements. Aimed at satisfying user requirements, it overcomes the gap between end-user requirements and agents' technical implementation. We first introduce the AgileGen framework, which is Agile-based for generative software development. This framework incorporates behavior-driven development, requiring end-users to make three decisions: Requirement Decision-making, Scenarios Decision-making, and Acceptance \& Recommendations, to develop reliable software applications through human-agent collaboration. We utilize the Gherkin language with acceptance criteria to describe user stories, expanding user requirements into different scenarios. We generate factors consistent with business logic from these testable user stories, which drives code generation. This factor ensures that end-user requirements align with the app's business logic while optimizing the end-user experience.
Moreover, we have developed AgileGen as a user-friendly interactive system. Due to its high scalability and generalizability design philosophy, we can extend our method to various human-collaborative tasks. In the future, we will incorporate appropriate external tools such as Cucumber to implement automated testing iterations. This will further enhance the consistency of the generated software with business logic.

\bibliographystyle{ACM-Reference-Format}
\bibliography{reference}

\end{document}